\documentclass[prd,aps,showpacs,tightenlines]{revtex4}  
\usepackage{mathrsfs}
\usepackage{amsmath}
\usepackage{amssymb}
\usepackage{epsfig}
\usepackage{graphicx}
\usepackage{booktabs}
\usepackage{multirow}
\usepackage{subfigure}

\begin{document}
\newcommand{\psl}{ p \hspace{-1.8truemm}/ }
\newcommand{\nsl}{ n \hspace{-2.2truemm}/ }
\newcommand{\vsl}{ v \hspace{-2.2truemm}/ }
\newcommand{\epsl}{\epsilon \hspace{-1.8truemm}/\,  }


\title{  $S$-wave contributions to the  $B_{(s)}\rightarrow \chi_{c1} (\pi\pi,K\pi,KK)$ decays}
\author{Meng-Kun Jia$^1$}
\author{Chao-Qi Zhang$^1$}
\author{Jia-Ming Li$^1$}
\author{Zhou Rui$^1$}\email{jindui1127@126.com}
\affiliation{$^1$College of Sciences, North China University of Science and Technology, Tangshan 063009,  China}

\date{\today}

\begin{abstract}
We make a detailed study of the three-body decays $B_{(s)}\rightarrow \chi_{c1} hh'$,
where $h^{(')}$ is either a pion or kaon,
by taking into account the $S$-wave states in the $hh'$ invariant mass distribution within the perturbative QCD approach.
The two meson distribution amplitudes are introduced  to capture
 the strong interaction related to the production of the $hh'$ system.
We calculate the branching ratios for the $S$-wave components and observe
 large values of order $10^{-4}$ for some Cabibbo-favored decays,
which are   accessible to the LHCb and Belle II experiments.
The obtained branching ratio $\mathcal{B}(B\rightarrow \chi_{c1}K^*_{0}(1430)(\rightarrow K^+\pi^-))=(5.1^{+0.6}_{-0.8})\times 10^{-5}$
consistent with the data from Belle within errors.
Moreover, we also predict the differential distributions in the $hh'$ invariant mass for the decays  under consideration,
which await the future experimental test.
 In addition, the corresponding $\chi_{c1}(2P)$ channels are also investigated, which are helpful to clarify the nature of the $X(3872)$ state.

\end{abstract}

\pacs{13.25.Hw, 12.38.Bx, 14.40.Nd }

\maketitle

\section{Introduction}
$B$ meson decays to final states containing a charmonium meson
have played a crucial role in the observation of $CP$ violation in the weak interactions of quarks
and provided powerful probes of the strong interaction in a heavy meson system.
In particular, the $\chi_{c1}$ modes,
which are allowed under the factorization hypothesis,
are found to be a comparison of production rates with respect to the similar $J/\psi$ processes~\cite{pdg2020}.
Studying the production of the $\chi_{c1}$ meson and its radial excited states in $B$ meson decays will help to
shed light on the production mechanisms in the exclusive charmonium $B$ decays.

Two-body decays of  $B\rightarrow \chi_{c1}\pi$~\cite{prd74051103,prd78091104}
and $B\rightarrow \chi_{c1} K^{(*)}$~\cite{plb634155,prl89011803,prl94141801} have been observed and well measured
by several  collaborations. Also, some multibody decay modes,
such as $B^0\rightarrow \chi_{c1}K^+\pi^-$~\cite{prd78072004},  $B_s\rightarrow \chi_{cJ}K^+K^-$~\cite{jhep081912018},
and $B\rightarrow \chi_{c1}\pi\pi K$~\cite{prd93052016}, were observed,
for which one can search for charmonium or charmoniumlike exotic states
in the pion-charmonium invariant mass distribution.
For example, the narrow exotic resonance  $X(3872)$ was discovered in the $J/\psi \pi^+\pi^-$ invariant mass spectrum
produced in $B\rightarrow J/\psi\pi^+\pi^-K$ decays by the Belle experiment~\cite{prl91262001},
and later confirmed by multiple other experiments~\cite{prl93072001,D0:2004zmu,BaBar:2004oro,LHCb:2011zzp}.
In addition, the  $X(3872)$ state was also observed in  $B\rightarrow X(3872)K\pi$ decays~\cite{prd91051101}.
Its quantum number assignment has been  identified to be $J^{PC}=1^{++}$~\cite{prl98132002,prd84052004,prl110222001},
suggesting it may be the typical $\chi_{c1}(2P)$ charmonium state in the quark model scenario.
However, its mass ($3871.69\pm0.17$ MeV), narrow width ($\Gamma<1.2$ MeV)~\cite{pdg2020}, and
the isospin violating decay chain $X(3872)\rightarrow J/\psi\rho^0 \rightarrow J/\psi \pi^+\pi^- $
\cite{prd84052004,jhep042013154,prl96102002,jhep012017117,prl98132002,prl110222001,prd92011102}
imply that it may not be a simple $c\bar c$ charmonium state.
Results from recent LHCb studies~\cite{prl126092001,prd102092005,jhep082020123} also support
it may have further mystery substructure beyond the conventional charmonium model.
Popular interpretations, including  $\chi_{c1}(2P)$ state, tetraquark, molecular state, admixture state, $c\bar c g$ hybrid state,
and vector glueball, have been proposed~\cite{explanations,Tornqvist:2004qy,Swanson:2003tb,Wong:2003xk,Maiani:2004vq,Li:2004sta,Seth:2004zb,Matheus:2009vq,Suzuki:2005ha,Kalashnikova:2005ui,Takizawa:2012hy,Chen:2013pya,
Wallbott:2019dng,Matheus:2006xi,Dubnicka:2010kz,Colangelo:2007ph,Butenschoen:2019npa,Coito:2012vf}, which means the question of its internal structure remains open.
A more detailed discussion of the current knowledge of the $X(3872)$ properties
can be found in Ref.~\cite{rmp90015003} and references therein.

$B$ meson  decays into final states containing the $\chi_{c1}(1P,2P)$ state
have generated many theoretical discussions
\cite{plb59191,plb568127,prd69054009,prd59054003,npb811155,prd71114008,epjc78463,prd87074035,epjc49643,Wang:2007fs}.
In particular, the authors of Ref.~\cite{prd87074035} analyzed the two-body $B\rightarrow \chi_{c1}(1P,2P)$ decays in QCD factorization
by treating charmonia as nonrelativistic bound states.
They found that the $B\rightarrow \chi_{c1}(2P)K$ decay rate can be comparable to that of the $\chi_{c1}(1P)$ mode
and argued that $X(3872)$ may be dominated by the $\chi_{c1}(2P)$ charmonium but mixed with a $D^0\bar D^{*0}/D^{*0}\bar D^0$ molecule state.
In Ref.~\cite{epjc49643},
 the branching ratio of  the $B\rightarrow X(3872)K$ was calculated in the perturbative QCD (PQCD) approach,
by assuming $X(3872)$ to be a regular $\chi_{c1}(2P)$ charmonium state.
The obtained number is larger than the current upper bound set by Belle~\cite{prd97012005} within the error bar,
which indicate  a pure charmonium assignment for $X(3872)$ is not suitable.
 Further studies should be carried out in other exclusive decays  to clarify its inner structure,
 especially in the three-body $B\rightarrow \chi_{c1}(2P)hh'$ decays,
 which are still lacking in the literature  to date.

In this consideration,
we study the three-body decays  $B\rightarrow \chi_{c1}hh'$ with $h,h'=\pi,K$ within the framework of PQCD,
where $\chi_{c1}$ is used to denote the $\chi_{c1}(1P)$ and $\chi_{c1}(2P)$  collectively.
The latter could help to clarify the nature of the  $X(3872)$ since $\chi_{c1}(2P)$
may be one of the possible assignments for $X(3872)$ as mentioned above.
Here we put the focus on  the $hh'$ pair originating from an $S$-wave configuration,
while the subjects related to the crossed channel such
as $\chi_{c1}h^{(')}$  and other higher partial wave
are outside the ambit of the present analysis.
For recent works applying triangle singularities to interpreting several charmoniumlike structures in the $X_{cc}\pi^+$ invariant mass distributions of $\bar{B}^0\rightarrow  X_{cc}K^-\pi^+$ with $X_{cc}= J/\psi,\psi(2S),\chi_{c1}$, we refer the reader to Refs.~\cite{Nakamura:2019emd,Nakamura:2019btl,Nakamura:2019nch}.

The PQCD approach has been successfully applied to various
three-body charmonium decays of $B$ meson to investigate the contributions of the resonances involved
\cite{prd91094024,epjc79792,prd97033006,prd98113003,prd99093007,prd101016015,cpc44073102,npb924745}.
The method has also been extended to the four-body charmless hadronic $B$ meson decays very recently~\cite{zjhep,Li:2021qiw}.
Within the quasi-two-body approximation,
we assume two light final-state mesons $h$ and $h'$  move almost in parallel for producing a resonance.
 The associated final-state interactions inside  $hh'$ pair are parametrized into 
the  nonperturbative  two meson distribution amplitudes (DAs)~\cite{G,G1,DM,Diehl:1998dk,Diehl:1998dk1,Diehl:1998dk2,MP}.
That is, three-body processes  are assumed to proceed predominantly via one
intermediate state which strongly decays into two light mesons.
The corresponding  decay amplitude  can be conceptually written as
the convolution of all the perturbative and nonperturbative objects:
\begin{eqnarray} \label{eq:fac}
\mathcal{A}=\Phi_B \otimes H\otimes \Phi_{hh'} \otimes \Phi_{\chi_{c1}},
\end{eqnarray}
where $\Phi_B$ and $\Phi_{\chi_{c1}}$ are the nonperturbative $B$ meson  and charmonium DAs, respectively.
The two meson DA $\Phi_{hh'}$   absorbs the nonperturbative
dynamics of the hadronization processes in the $hh'$ system.
The hard kernel $H$, similar to the case of two-body decays, includes the leading-order contributions
plus the vertex corrections. As pointed out in Refs.~\cite{npb811155,prd87074035},
the infrared divergences arising from vertex corrections cancel
in the $B\rightarrow \chi_{c1}$ decay as in the case of $B\rightarrow J/\psi$.
Therefore, the vertex corrections obtained in QCDF can be applied to PQCD without introducing any extra parton transverse momenta~\cite{prd98113003}.

The paper is organized as follows. After the Introduction,
we present our model kinematics and describe the $S$-wave DAs in $\pi\pi$, $K\pi$, and $KK$ pairs, respectively.
In Sec.~\ref{sec:results}, we make predictions of the branching ratio and the differential distribution for each $S$-wave component
in the considered three-body decays.
 In the final section, we give discussions and the conclusion.
Some technical details are relegated to the Appendix.

\section{Kinematics and the $S$-wave two meson distribution amplitudes  }\label{sec:framework}
\begin{figure}[!htbh]
\begin{center}
\vspace{1.5cm} \centerline{\epsfxsize=7cm \epsffile{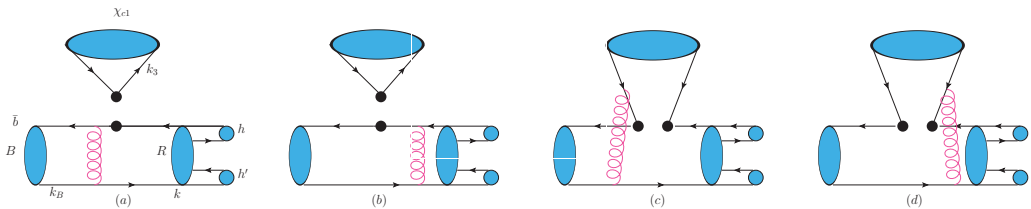}}
\vspace{1.0cm}
\caption{Feynman diagrams for the $B_{(s)}\rightarrow \chi_{c1} R(\rightarrow hh')$ decays  at
the leading-order approximation, where the symbol $\bullet$   denotes the
insertion of effective weak interaction.}
 \label{fig:femy}
\end{center}
\end{figure}
Consider the quasi-two-body process $B_{(s)}\rightarrow \chi_{c1}R(\rightarrow hh')$,
whose leading order  diagrams are shown in Fig.~\ref{fig:femy}.
In the rest frame of the $B_{(s)}$ meson,
we assume the final state charmonium is moving along the direction of $v=(0,1,\textbf{0}_{\text{T}})$
while the meson pair is along $n=(1,0,\textbf{0}_{\text{T}})$, where $n$ and $v$ are two lightlike vectors in the light-cone coordinates.
Then the $B_{(s)}$ meson momentum $p_B$, the $\chi_{c1}$ meson momentum $p_3$,
and the meson pair momentum $p$ can be parametrized as~\cite{210503899}
\begin{eqnarray}
 p_B&=&\frac{M}{\sqrt{2}}(1,1,\textbf{0}_{T}),
 \quad p_3=\frac{M}{\sqrt{2}}(g^-,g^+,\textbf{0}_{T}),
 \quad  p=\frac{M}{\sqrt{2}}(f^+,f^-,\textbf{0}_{T}),
\end{eqnarray}
where the variables
\begin{eqnarray}
f^{\pm}&=&\frac{1}{2}(1+\eta-r^2\pm\sqrt{(1-\eta)^{2}-2r^2(1+\eta)+r^{4}}),\nonumber\\
g^{\pm}&=&\frac{1}{2}(1-\eta+r^2\pm\sqrt{(1-\eta)^{2}-2r^2(1+\eta)+r^{4}}),
\end{eqnarray}
with the mass ratio $r=m/M$ and $m(M)$ is the mass of the charmonium ($B$ meson).
The factor  $\eta$ is defined as $\eta=\omega^2/M^2$ with $\omega$ being the invariant mass of the meson pair satisfying $p^2=\omega^2$.
The meson momenta $p_{1}$ and $p_{2}$ inside meson pair,
obeying  momentum conservation $p=p_1+p_2$ and the on-shell conditions $p_{1,2}^2=m_{1,2}^2$,
one can derive them:
\begin{eqnarray}\label{eq:p1p2}
p_1&=&\left(\frac{M}{\sqrt{2}}(\zeta+\frac{r_1-r_2}{2\eta})f^+,\frac{M}{\sqrt{2}}(1-\zeta+\frac{r_1-r_2}{2\eta})f^-,\textbf{p}_{T}\right),\nonumber\\
p_2&=&\left(\frac{M}{\sqrt{2}}(1-\zeta-\frac{r_1-r_2}{2\eta})f^+,\frac{M}{\sqrt{2}}(\zeta-\frac{r_1-r_2}{2\eta})f^-,-\textbf{p}_{T}\right),
\end{eqnarray}
with the mass ratios $r_{1,2}=m^2_{1,2}/M^2$. $\zeta$ is the meson momentum fraction up to corrections from the meson masses~\cite{zjhep}.
By use of the on-shell conditions $p^2_{1,2}=m^2_{1,2}$,
the transverse momentum $\textbf{p}_{T}$ can be written as
\begin{eqnarray}
|\textbf{p}_{T}|^2=\omega^2[\zeta(1-\zeta)+\frac{(r_{1}-r_{2})^2}{4\eta^2}-\frac{r_{1}+r_{2}}{2\eta}].
\end{eqnarray}
In the $hh'$ rest frame,  the three-momenta of the final states $h$ and charmonium are written as
\begin{eqnarray}
|\vec{p}_1|=\frac{\sqrt{\lambda(\omega^2,m_1^2,m_{2}^2)}}{2\omega}, \quad
|\vec{p}_3|=\frac{\sqrt{\lambda(M^2,m^2,\omega^2)}}{2\omega},
\end{eqnarray}
respectively, with the K$\rm\ddot{a}$ll$\rm\acute{e}$n function $\lambda (a,b,c)= a^2+b^2+c^2-2(ab+ac+bc)$.
To evaluate the hard kernels, the following parametrization for the
valence quark momenta labeled by $k_B$, $k_3$, and $k$ in Fig.~\ref{fig:femy} is useful:
\begin{eqnarray}\label{eq:kt}
  k_B&=&(0,\frac{M}{\sqrt{2}}x_B,\textbf{k}_{BT}),\quad
  k_3=(\frac{M}{\sqrt{2}}g^-x_3,\frac{M}{\sqrt{2}}g^+x_3,\textbf{k}_{3T}),\quad
  k=(\frac{M}{\sqrt{2}}f^+z,0,\textbf{k}_{T}),
\end{eqnarray}
with the parton momentum fractions $x_B,x_3,z$ and the corresponding transverse momenta $\textbf{k}_{BT}, \textbf{k}_{3T}, \textbf{k}_T$.

The light-cone hadronic matrix element for a $B_{(s)}$ meson is decomposed as \cite{ppnp5185}
\begin{eqnarray}\label{b}
\Phi_{B_{(s)}}(x,b)=\frac{i}{\sqrt{2N_c}}[(\rlap{/}{p_B}+M)\gamma_5\phi_{B_{(s)}}(x,b)],
\end{eqnarray}
with $b$ being the conjugate variable of the parton transverse momentum $k_{T}$, and $N_c$ denoting the number of colors.
We here only consider the leading Lorentz structure,
while other subleading contributions~\cite{Li:2014xda,Li:2012nk} are negligible within the accuracy of the current work.
For the $B_{(s)}$ meson DAs, we adopt the conventional form~\cite{ppnp5185,prd65014007},
\begin{eqnarray}
\phi_{B_{(s)}}(x,b)=N_{B_{(s)}} x^2(1-x)^2\exp\left[-\frac{x^2M^2}{2\omega^2_{B_{(s)}}}-\frac{\omega^2_{B_{(s)}}b^2}{2}\right],
\end{eqnarray}
with the shape parameter $\omega_{B}=0.40\pm 0.04$ GeV for $B_{u,d}$ mesons and
$\omega_{B_{s}}=0.48 \pm 0.05$ GeV for a $B_s$ meson~\cite{201215074}. The normalization constant
$N_{B_{(s)}}$ is related to the $B_{(s)}$ meson decay constant $f_{B_{(s)}}$ via the normalization
\begin{eqnarray}
\int_0^1\phi_{B_{(s)}}(x,b=0)d x=\frac{f_{B_{(s)}}}{2\sqrt{2N_c}}.
\end{eqnarray}
For more alternative models of $B$ meson DA, one can refer to~\cite{prd102011502,prd70074030,Li:2012md}.

The distribution amplitudes of $\chi_{c1}$,
defined via the nonlocal matrix element, have been derived in Ref~\cite{prd97033001}.
The longitudinal polarization component is given by
\begin{eqnarray}
\Phi _{\chi_{c1}}=\frac{1}{{\sqrt{2 N_{c}}}}\gamma_{5}\rlap{/}{\epsilon _L}(m \chi _{L}(x)+\chi _{t}(x) \rlap{/}{p_{3}}),
\end{eqnarray}
with the longitudinal polarization vector $\epsilon _L=\frac{1}{\sqrt{2}\eta}(-g^-,g^+,\textbf{0}_{T})$.
The twist-2 and  twist-3 DAs are  collected  as follows:
\begin{eqnarray}
\chi _{L}(x)&=&\frac{f_{\chi_{c1}}}{2\sqrt{6}}N_{L}x(1-x)\mathcal{T}(x), \nonumber\\
\chi _{t}(x)&=&\frac{f_{\chi_{c1}}^{\bot}}{2\sqrt{6}}\frac{N_{T}}{6}(2x-1)[1-6x+6x^{2}]\mathcal{T}(x),
\end{eqnarray}
 where $f^{(\bot)}_{\chi_{c1}}$ is the vector (tensor) decay constants.
The coefficients $N_{L,T}$ satisfy the normalization conditions~\cite{prd97033001,Wang:2013ywc}
\begin{eqnarray}
\int_0^1N_Lx(1-x)\mathcal {T}(x)dx=1, \quad
 \int_0^1N_Tx(1-x)(2x-1)^2\mathcal {T}(x)dx=1.
\end{eqnarray}
The function $\mathcal{T}(x)$ can be extracted from $P$-wave Schr$\rm\ddot{o}$dinger states for a Coulomb potential.
The explicit expression for $1P$ state can be found in Refs.~\cite{prd97033001,Li:2020app},
while that of the $2P$ state will be derived in the Appendix.

The light-cone matrix element for an $S$-wave meson pair is decomposed, up to the twist 3, into~\cite{plb561258,prd91094024}
\begin{eqnarray}\label{eq:fuliye2}
\Phi_{hh'}^{S-wave}=\frac{1}{\sqrt{2N_c}}[\rlap{/}{P}\phi^0_S(z,\omega)+
\omega\phi^s_S(z,\omega)+\omega(\rlap{/}{n}\rlap{/}{v}-1)\phi^t_S(z,\omega)]P_l(2\zeta-1),
\end{eqnarray}
where the Legendre polynomials $P_l(2\zeta-1)=1$ for the $S$-wave component.
The two-meson DAs are parametrized as
\begin{eqnarray}\label{eq:phi0st}
\phi^0_S(z,\omega)&=&
\left\{
\begin{aligned}
&\frac{9 F_{hh'}(\omega)}{\sqrt{2N_c}}a_{hh'}z(1-z)(1-2z),   &hh'=\pi\pi,KK, \\
&\frac{3 F_{hh'}(\omega)}{\sqrt{2N_c}}z(1-z) \left[\frac{1}{\mu_S}+B_13(1-2z)
+B_3 \frac{5}{2}(1-2z)(7(1-2z)^2-3)\right],  &hh'=K\pi, \\
\end{aligned}\right. \nonumber\\
\phi^s_S(z,\omega)&=&\frac{ F_{hh'}(\omega)}{2\sqrt{2N_c}},\nonumber\\
\phi^t_S(z,\omega)&=&\frac{ F_{hh'}(\omega)}{2\sqrt{2N_c}}(1-2z),
\end{eqnarray}
with the Gegenbauer moments $a_{\pi\pi}=0.2\pm 0.2$~\cite{prd91094024}, $a_{KK}=0.80\pm0.16$~\cite{epjc79792},
$B_1=-0.57\pm 0.13$ and $B_3=-0.42\pm 0.22$~\cite{prd73014017,prd77014034}.
Except for the twist-2 DA of the $K\pi$ pair, others are set to the asymptotic forms
since the coefficients in the Gegenbauer expansion of the two meson DAs are poorly known at the moment.

The timelike scalar form factors $F_{hh'}(\omega)$,
 which absorbs the elastic rescattering effects in a final-state meson pair,
 are parametrized in different forms according to different meson pairs.
 Let us begin with the   pion pair first.
 There are two intriguing light scalar resonances $f_0(500)$ and $f_0(980)$ below or near 1 GeV, which can couple to $\pi^+\pi^-$.
Their internal structures were quite controversial~\cite{Stone:2013eaa,Fleischer:2011au,Jaffe:2004ph,Klempt:2007cp}.
The LHCb collaboration observed a peak for the $f_0(980)$ in the $B^0_s\rightarrow J/\psi \pi^+\pi^-$ decay,
while that of $f_0(500)$  is not seen~\cite{Aaij:2014emv}.
On the contrary, in the corresponding $B_s$ decay,  a signal is seen for the $f_0(500)$ production, but no visible trace for $f_0(980)$ production.
Therefore, in this work, we assume
the $f_0(500)$ and $f_0(980)$ enter into the nonstrange and strange scalar form factors, respectively.
The former  contribute dominantly in the  $B^0$ decay,
while the latter should feature mainly in the $B^0_{s}$ mode.

Including  higher resonances,
the strange scalar form factor
 can be described as the coherent sum of three scalar resonances $f_0(980)$, $f_0(1500)$, and $f_0(1790)$,
which have been widely employed in the PQCD studies of
the $B_s\rightarrow (J/\psi,\psi(2S), \eta_c,\eta_c(2S))\pi\pi$ decays~\cite{prd91094024,cpc41083105,epjc76675,epjc77199}.
Explicitly, we have~\cite{prd91094024}
\begin{eqnarray}
F_{\pi\pi}^{s\bar{s}}(\omega)&=& \frac{c_{1}m_{f_{0}(980)}^{2}e^{i\theta_{1}}}{m_{f_{0}(980)}^{2}
-\omega^{2}-im_{f_{0}(980)}(g_{\pi\pi}\rho_{\pi\pi}+g_{KK}\rho_{KK})}\nonumber\\
 && +\frac{c_{2}m_{f_{0}(1500)}^{2}e^{i\theta_{2}}}{m_{f_{0}(1500)}^{2}-\omega^{2}-im_{f_{0}(1500)}\Gamma_{f_{0}(1500)}(\omega^{2})}\nonumber\\
 && +\frac{c_{3}m_{f_{0}(1790)}^{2}e^{i\theta_{3}}}{m_{f_{0}(1790)}^{2}-\omega^{2}-im_{f_{0}(1790)}\Gamma_{f_{0}(1790)}(\omega^{2})},
\end{eqnarray}
where $c_i$ and $\theta_i$, $i=1,2,3$, are the corresponding  weight coefficients and phases of the resonances
with the parameter values of Ref.~\cite{prd91094024}.
$m_{f_0}$ is the nominal mass of the resonance.
 $\Gamma_{f_0}(\omega)$, the mass-dependent width,
is defined as in the case of a scalar resonance
\begin{eqnarray}
\Gamma_{f_0}(\omega)=\Gamma_{f_0}\frac{m_{f_0}}{\omega}(\frac{\omega^{2}-4m_{\pi}^{2}}{m_{f_0}^{2}-4m_{\pi}^{2}})^{\frac{1}{2}}, 
\end{eqnarray}
where $\Gamma_{f_0}$ is the partial width of the resonance.
Different from $f_0(1500)$ and $f_0(1790)$ resonances  described usually by the Breit-Wigner (BW) model,
$f_0(980)$ is parametrized as the Flatt\'{e} model~\cite{plb63228} since its mass is close to
the $K\bar K$ threshold.
The constants $g_{\pi\pi}$ and $g_{KK}$ are the $f_0(980)$ couplings to $\pi^+\pi^-$ and $K^+K^-$ final states, respectively.
The $\rho_{\pi\pi(KK)}$ factors are given by the Lorentz-invariant phase space
\begin{eqnarray}
\rho_{\pi\pi}&=&\frac{2}{3}\sqrt{1-\frac{4m_{\pi^{\pm}}^{2}}{\omega^{2}}}+\frac{1}{3}\sqrt{1-\frac{4m_{\pi^{0}}^{2}}{\omega^{2}}},\nonumber\\
\rho_{KK}&=&\frac{1}{2}\sqrt{1-\frac{4m_{K^{\pm}}^{2}}{\omega^{2}}}+\frac{1}{2}\sqrt{1-\frac{4m_{K^{0}}^{2}}{\omega^{2}}}.
\end{eqnarray}

 For the nonstrange case,  only the resonance $f_{0}(500)$ is included here,
which can be modeled using two alternative approaches, the BW function~\cite{prd91094024} and the Bugg formula~\cite{jpg34151}.
The BW model read as
\begin{eqnarray}
F_{\pi\pi}^{d\bar{d}}(\omega)=\frac{c_{BW}m_{f_{0}(500)}^{2}}{m_{f_{0}(500)}^{2}-\omega^{2}-im_{f_{0}(500)}\Gamma_{f_{0}(500)}(\omega^{2})},
\end{eqnarray}
with $c_{BW}=3.5$~\cite{prd91094024}.
The Bugg resonant line shape~\cite{jpg34151}, with more theoretically motivated shape parameters,
  has been used by several recent   analyses, e.g. Refs.~\cite{epjc76675,cpc41083105,Aaij:2015sqa,Aaij:2014vda},
\begin{eqnarray}\label{eq:bugg}
F_{\pi\pi}^{d\bar{d}}(\omega)=c_{Bugg}m_r\Gamma_{1}(\omega^2)[m_r^{2}-\omega^2-g_{1}^{2}\frac{\omega^2-s_{A}}{m_r^{2}-s_{A}}
z(\omega^2)-im_r\sum_{i=1}^{4}\Gamma_{i}(\omega^2)]^{-1},
\end{eqnarray}
with $c_{Bugg} $ being a tunable parameter.
Its value is set to 1.6 so that the corresponding PQCD prediction of $\mathcal{B} (\bar{B}^0\rightarrow J/\psi f_0(500)(\rightarrow \pi^+\pi^-))=8.5\times 10^{-6}$
is consistent with the LHCb data $(8.8\pm0.5^{+1.1}_{-1.5})\times 10^{-6}$ in Ref.~\cite{prd90012003}.
Variables and parameters in the above equation are not shown here for readability, which can be found in~\cite{jpg34151}.
We also note that the $f_{0}(500)$ can be represented as a simple pole~\cite{Pelaez:2015qba,Oller:2004xm,LHCb:2019sus,Cheng:2020iwk},
parametrized as
\begin{eqnarray}\label{eq:pole}
A_{\sigma}(\omega)=\frac{1}{\omega^2-s_{\sigma}},
\end{eqnarray}
where $s_{\sigma}$ is the square of the pole position $\sqrt{s_{\sigma}}=m_{\sigma}-i\Gamma_{\sigma}$, extracted from the data.
However, Eq.~(\ref{eq:pole}) carries a dimension and is not normalized to the unity as $\omega \rightarrow 0$.
Thus it is not appropriate to  be taken as a form factor in the PQCD approach in the current form.
The exact form factor corresponding to the pole model in PQCD should be taken into account in the future.

For the scalar form factor of the $K\pi$ system, we employ the LASS line shape~\cite{npb296493},
which consists of the $K_0^*(1430)$ resonance as well as an effective-range nonresonant component,
\begin{eqnarray}\label{eq:sform}
F_{K\pi}(\omega)&=&\frac{\omega}{|\vec{p}_1|}\cdot\frac{1}{\cot \delta_B-i}+
e^{2i \delta_B}\frac{m_0^2\Gamma_0/|\vec{p}_0|}{m_0^2-\omega^2-im_0^2
\frac{\Gamma_0}{\omega}\frac{|\vec{p}_1|}{|\vec{p}_0|}}, \nonumber\\
\cot\delta_B&=&\frac{1}{a|\vec{p}_1|}+\frac{1}{2}b|\vec{p}_1|.
\end{eqnarray}
$m_0$ and $\Gamma_0$ are the pole mass and width
of the $K^*_{0}(1430)$, while the scattering lengths $a$ and $b$ effective range are parameters that describe the shape,
whose numbers are taken from measurements  at the
LASS experiment  and tabulated in the next section.
$\vec{p}_0$ is the value of $\vec{p}_1$ calculated using the nominal resonance mass, $m_0$.
The phase factor $\delta_B$ is needed for the conservation of unitarity.
It is worth noting that the LASS parametrization has a range of applicability
up to about the charm hadron mass~\cite{Aubert:2005ce,BaBar:2005qms} in the $K\pi$ invariant mass,
which is just approaching the upper bound of $M-m$ for the charmonium $B$ decays.
Therefore, it is not necessary to introduce a nonphysical cutoff here and
 the  LASS model is  appropriate to describe  the $K\pi$ $S$-wave in the decays under study.

As for the case of the $K\bar K$ system, we follow Ref.~\cite{epjc79792}  to take the form as,
\begin{eqnarray}
F_{KK}(\omega)&=&[\frac{m^2_{f_0(980)}}{m^2_{f_0(980)}-\omega^2-i m_{f_0(980)}(g_{\pi\pi}\rho_{\pi\pi}+g_{KK}\rho_{KK}F_{KK}^2)}\nonumber\\&&
+\frac{c_{f_0(1370)}m_{f_{0}(1370)}^{2}}{m_{f_{0}(1370)}^{2}-\omega^{2}-im_{f_{0}(1370)}\Gamma_{f_{0}(1370)}(\omega^{2})}](1+c_{f_0(1370)})^{-1}.
\end{eqnarray}
with $c_{f_0(1370)}=0.12e^{-i\pi/2}$~\cite{epjc79792}, which yields the branching ratios
of the $f_0(980)$ and $f_0(1370)$ components in $B_s\rightarrow J/\psi K^+K^-$  are
 consistent with the data, simultaneously.
The exponential term $F_{KK}=e^{-\alpha q_K^2}$
is introduced above the $KK$ threshold to reduce the $\rho_{KK}$ factor as $\omega$ increases,
where $q_k$ is the momentum of the kaon in the $KK$ rest frame and $\alpha=2.0\pm 0.25$ GeV$^{-2}$~\cite{Aaij:2014emv,Bugg:2008ig}.

Now we will present the formulas of  amplitude for the quasi-two-body decay mediated by scalar resonances.
Performing the calculations to the factorizable and nonfactorizable diagrams in Fig.~\ref{fig:femy},
one gets the following expressions:
\begin{eqnarray}
\mathcal{F}^{LL}&=&-8\pi C_F f_{\chi_{c1}} M^3 \int_0^1dx_B dz \int_0^{\infty} b_B db_Bbdb\phi_B(x_B,b_B)\nonumber\\&&
\{[\omega  \phi^s_S \left(-2 f^+ g^+ z-g^-+g^+\right)+\omega  \phi^t_S \left(-2 f^+ g^+ z+g^-+g^+\right)+M \phi^0_S \left(f^+ g^+ \left(f^+ z+1\right)-f^- g^-\right)]\nonumber\\
&&\alpha_s(t_a)e^{-S_{ab}(t_a)}h(\alpha_e,\beta_a,b_B,b_1)S_t(x_3)+[M \phi^0_S  \left(\left(f^-+1\right) f^+ g^+-g^- \left(f^-+f^+ \left(f^--x_{B}\right)\right)\right)-2 \omega  \phi^s_S\nonumber\\
&& \left(g^- \left(-f^-+x_{B}-1\right)+\left(f^++1\right) g^+\right)]\alpha_s(t_b)e^{-S_{ab}(t_b)}h(\alpha_e,\beta_b,b_1,b_B)S_t(x_B)\},
\end{eqnarray}
\begin{eqnarray}
\mathcal{M}^{LL}&=&16\sqrt{2}\pi C_F  M^3 \int_0^1dx_B dz dx_3\int_0^{\infty} b_B db_Bb_3db_3\phi_B(x_B,b_B)
\nonumber\\&&[M r \chi _{L}(x_3) \phi^0_S \left(2 f^+ \left(g^+\right)^2x_{3}+g^-
\left(f^+ x_{B}+f^- \left(-2 f^+ z-2 g^- x_{3}+x_{B}\right)\right)+f^+ g^+ \left(\left(f^-+f^+\right) z-2x_{B}\right)\right)\nonumber\\
&&-2 r \chi _{L}(x_3) \omega \phi^s_S \left(f^+ g^+ z+g^- x_{B}\right)+4 g^- g^+ r_{c} \chi _{t}(x_3) \omega \phi^t_S]
\alpha_s(t_d)e^{-S_{cd}(t_d)}h(\beta_d,\alpha_e,b,b_B),
\end{eqnarray}
\begin{eqnarray}
\mathcal{F}^{LR}=-\mathcal{F}^{LL}, \quad \mathcal{M}^{SP}=\mathcal{M}^{LL},
\end{eqnarray}
with $r_c=m_c/M$ and $m_c$ is the charm quark mass;  $C_f=4/3$ is a color factor.
 The superscripts $LL$, $LR$, and $SP$  refer to the contributions from $(V-A)\otimes(V-A)$, $(V-A)\otimes(V+A)$
and $(S-P)\otimes(S+P)$ operators, respectively.
The hard functions $h$ and the threshold resummation factor $S_t(x)$ are adopted from Ref.~\cite{epjc77610}.
$\alpha$ and $\beta_i$ with $i=a,b,d$ denote the virtuality of the internal gluon and quark, respectively,
expressed as
\begin{eqnarray}\label{eq:mm}
\alpha &=&zx_Bf^+M^2, \nonumber\\
\beta_a&=&zf^+M^2,\nonumber\\
\beta_b&=&-f^+(f^--x_B)M^2,\nonumber\\
\beta_d&=&-(f^+z+g^-x_3)(g^+x_3-x_B)M^2+m_c^2.
\end{eqnarray}
The hard scale $t_i$ is chosen as the largest scale of the virtualities of the internal particles
in the hard amplitudes:
\begin{eqnarray}\label{eq:scale}
t_a&=&\max(\sqrt{\alpha},\sqrt{\beta_a},1/b,1/b_B),\nonumber\\
t_b&=&\max(\sqrt{\alpha},\sqrt{\beta_b},1/b,1/b_B),\nonumber\\
t_d&=&\max(\sqrt{\alpha},\sqrt{\beta_d},1/b_3,1/b_B).
\end{eqnarray}
The Sudakov factors derived with the leading-logarithm $k_T$ resummation are given by
\begin{eqnarray}\label{eq:ss}
S_{ab}(t)&=&s(\frac{M}{\sqrt{2}}x_B,b_B)+s(\frac{M}{\sqrt{2}}f^+z,b)+s(\frac{M}{\sqrt{2}}f^+(1-z),b)
+\frac{5}{3}\int_{1/b_B}^t\frac{d\mu}{\mu}\gamma_q(\mu)+2\int_{1/b}^t\frac{d\mu}{\mu}\gamma_q(\mu),\nonumber\\
S_{cd}(t)&=&s(\frac{M}{\sqrt{2}}x_B,b_B)+s(\frac{M}{\sqrt{2}}f^+z,b_B)+s(\frac{M}{\sqrt{2}}f^+(1-z),b_B)
+s(\frac{M}{\sqrt{2}}g^+x_3,b_3)+s(\frac{M}{\sqrt{2}}g^+(1-x_3),b_3)\nonumber\\&&
+\frac{11}{3}\int_{1/b_B}^t\frac{d\mu}{\mu}\gamma_q(\mu)+2\int_{1/b_3}^t\frac{d\mu}{\mu}\gamma_q(\mu),
\end{eqnarray}
where $\gamma_q=-\alpha_s/\pi$ is the quark anomalous dimension, and
the explicit expression of the function $s(Q,b)$ can be found in~\cite{epjc11695}.
We note that the complete next-to-leading-logarithm (NLL) $k_T$ resummation for the $J/\psi$ meson wave function
has been developed recently~\cite{Liu:2020upy}, which could be applicable to the $B\rightarrow \chi_{c1}$ decays.
However, it is found that its effect on the branching ratio is numerically small, less than $20\%$~\cite{Liu:2020upy};
thus its contribution is neglected  in subsequent calculations.



By combining the contributions from different diagrams
with the corresponding Wilson coefficients,
the full decay amplitude  can be recast to
\begin{eqnarray}
\mathcal{A}&=&\frac{G_F}{\sqrt{2}}\Big\{\lambda_{c}
\Big [a_2\mathcal{F}^{LL}+C_2\mathcal{M}^{LL} \Big]
-\lambda_{t}\Big [(a_3+a_9)\mathcal{F}^{LL}+\nonumber\\
&&(a_5+a_7)\mathcal{F}^{LR}
+(C_4+C_{10})\mathcal{M}^{LL}+(C_6+C_8)\mathcal{M}^{SP}\Big ]\Big\},
\end{eqnarray}
with
\begin{eqnarray}\label{eq:ai}
a_2=C_1+\frac{1}{3}C_2,
a_3=C_3+\frac{1}{3}C_4,
a_9=C_9+\frac{1}{3}C_{10},
a_5=C_5+\frac{1}{3}C_6,
a_7=C_7+\frac{1}{3}C_{8}.
\end{eqnarray}
 The quantities $\lambda_{p}\equiv V^*_{pb}V_{pq}$ with $p=c,t$ and $q=d,s$  encode the Cabibbo-Kobayashi-Maskawa (CKM) factors.
We also consider the vertex corrections,
whose effects can be combined into the  coefficients $a_i$ in Eq. (\ref{eq:ai}) as  \cite{qcdf,Beneke:2000ry,Beneke:2001ev}
\begin{eqnarray}\label{eq:vertex}
a_2&\rightarrow & a_2 +\frac{\alpha_s}{4\pi}\frac{C_f}{N_c}C_2
 \left[-18-12\text{ln}(\frac{t}{m_b})+f_I\right ] ,\nonumber\\ a_3+a_9 &\rightarrow &a_3+a_9
+\frac{\alpha_s}{4\pi}\frac{C_f}{N_c}(C_4+C_{10}) \left[-18-12\text{ln}(\frac{t}{m_b})+f_I\right ],\nonumber\\
a_5+a_7  &\rightarrow &a_5+a_7-\frac{\alpha_s}{4\pi}\frac{C_f}{N_c}(C_6+C_{8}) \left[-6-12\text{ln}(\frac{t}{m_b})+f_I\right ],
\end{eqnarray}
where the detail calculations  for $f_I$ refer to Refs. \cite{prd87074035,npb811155}.

Finally, the differential branching ratio for the  $B\rightarrow \chi_{c1}hh'$ process  reads as
\begin{eqnarray}\label{eq:dfenzhibi}
\frac{d \mathcal{B}}{d \omega}=\frac{\tau \omega|\vec{p}_1||\vec{p}_3|}{32\pi^3M^3}|\mathcal{A}|^2.
\end{eqnarray}

\section{RESULTS AND DISCUSSIONS}\label{sec:results}
This section serves to summarize all parameter values required for numerical calculations.
The meson and heavy quark masses (GeV), lifetimes (ps), and the Wolfenstein parameters are taken from Particle Data Group~\cite{pdg2020},
\begin{eqnarray}
M_{B_s}&=&5.37, \quad M_B=5.28,  \quad m_{K}=0.494, \quad m_{\pi}=0.14,\nonumber\\
m_{\chi_{c1}(1P)}&=&3.51067,\quad m_{\chi_{c1}(2P)}=3.87169, \quad {m}_b=4.8, \quad  \bar{m}_c(\bar{m}_c)=1.275,\nonumber\\
\tau_{B^+}&=&1.638,\quad  \tau_{B_s}=1.51, \quad \tau_{B^0}=1.51, \nonumber\\
\lambda &=& 0.22650, \quad  A=0.790,  \quad \bar{\rho}=0.141, \quad \bar{\eta}=0.357.
\end{eqnarray}
The  $B$ decay constants are set to the values $f_B=0.19$ GeV and  $f_{B_s}=0.24 $ GeV~\cite{zjhep}.
Since there is no measurement for  $f_{\chi_{c1}(2P)}$,
we assume $f^{(\perp)}_{\chi_{c1}(2P)}=f^{(\perp)}_{\chi_{c1}(1P)}=0.335$ GeV~\cite{prd71114008,epjc49643,epjc78463}
and do not distinguish the vector and tensor decay constants in subsequent calculations.
The relevant resonance parameters are listed in Table~\ref{tab:rp}.
Other parameters appearing in the  two meson DAs have been specified before.

Using the above parameters, we calculate the $CP$ average branching ratios of various $S$-wave components for the neutral $B^0$ and $B^0_s$ decays
by integrating the differential branching ratio  in Eq.~(\ref{eq:dfenzhibi}) with respect to $\omega$.
The corresponding numbers for the charge analogous $B^+$ decay processes can be obtained
by multiplying the $B^0$ ones with a factor of $\tau_{B^+}/\tau_{B^0}$ in the limit of isospin symmetry.
The theoretical errors correspond to the uncertainties due to
the shape parameters $\omega_{B_{(s)}}=0.40\pm 0.04(0.48\pm 0.05)$ GeV for the $B_{(s)}$ meson wave function,
the hard scales $t$ defined in Eq.~(\ref{eq:scale}),
which vary from $0.75t$ to $1.25t$, and  the Gegenbauer moments $a_{\pi\pi}=0.2\pm 0.2$~\cite{prd91094024}, $a_{KK}=0.80\pm0.16$~\cite{epjc79792},
$B_1=-0.57\pm 0.13$ and $B_3=-0.42\pm 0.22$~\cite{prd73014017,prd77014034}
associated with the twist-2 DAs as shown in Eq.~(\ref{eq:phi0st}), respectively.
It is necessary to stress that the twist-3 DAs of the meson pair in this work  are taken as the asymptotic forms for lack
of better results from nonperturbative methods, which may give significant uncertainties.
In the following, we will discuss the relevant numerical results in turn.
\begin{table}
\caption{A summary of the relevant resonance parameters.}
\label{tab:rp}
\begin{tabular}[t]{lcc}
\hline\hline
Resonance & Model& Parameters \\ \hline
$f_{0}(500) $  & $\text{BW}$ & $m_{R}=0.50$\text{GeV}, $\Gamma_{R}=0.40$\text{GeV} \cite{prd91094024}\\
& $\text{Bugg}$    &See Ref.~\cite{jpg34151}\\
$f_{0}(980)$   & $\text{Flatt$\acute{e}$}$    & $m_{R}=0.99$ \text{GeV}\cite{pdg2020}, $g_{\pi\pi}=0.167$\text{GeV}, $g_{KK}/g_{\pi\pi}=3.47$\cite{prd90012003}\\
$f_{0}(1370)$  & $\text{BW}$  & $m_{R}=1.475$\text{GeV}, $\Gamma_{R}=0.113$\text{GeV}\cite{Aaij:2014emv} \\
$f_{0}(1500)$  & $\text{BW}$ & $m_{R}=1.50$\text{GeV}, $\Gamma_{R}=0.12$\text{GeV}\cite{Aaij:2014emv}\\
$f_{0}(1790)$  & $\text{BW}$ & $m_{R}=1.81$\text{GeV}, $\Gamma_{R}=0.32$\text{GeV} \cite{prd91094024}\\
$(K\pi)_{S-\text{wave}}$ &$\text{LASS}$  \quad & $m_{R}=1.435$\text{GeV}, $\Gamma_{R}=0.279$\text{GeV}, $a=1.94 \text{GeV}^{-1}$,  $b=1.76 \text{GeV}^{-1}$\cite{BaBar:2008bxw}\\
\hline\hline
\end{tabular}
\end{table}
\subsection{ $B_s^0\rightarrow \chi_{c1}\pi^+\pi^-$}

\begin{table}
\caption{ Branching ratios  of $S$-wave resonant contributions to  the $B^0_s\rightarrow \chi_{c1}(1P,2P) \pi^+\pi^-$ decays.
The theoretical errors correspond to the uncertainties due to the shape parameters $\omega_{B_s}$, the hard scale $t$, and  the Gegenbauer moment $a_{\pi\pi}$, respectively.}
\label{tab:brpipi}
\begin{tabular}[t]{lcccc}
\hline\hline
Modes & $\mathcal{B}(R=f_{0}(980))$ & $\mathcal{B}(R=f_{0}(1500))$ & $\mathcal{B}(R=f_{0}(1790))$  & $\mathcal{B}(\text{S-wave})$\\ \hline
$B^0_s\rightarrow \chi_{c1}(1P) \pi^+\pi^- $  & $(7.6^{+3.4+0.5+0.9}_{-2.0-0.5-0.6})\times 10^{-5}$
& $(7.8^{+0.7+0.6+0.2}_{-0.9-0.4-0.3})\times 10^{-6}$  & $(6.4^{+1.5+0.0+0.5}_{-1.2-0.2-0.5})\times 10^{-7}$ & $(1.1^{+0.3+0.0+0.1}_{-0.3-0.1-0.1})\times 10^{-4}$ \\
$B^0_s\rightarrow \chi_{c1}(2P) \pi^+\pi^-$  & $(7.1^{+2.4+0.2+0.6}_{-1.8-0.5-0.4})\times 10^{-5}$ & $(1.8^{+0.5+0.0+0.2}_{-0.3-0.0-0.1})\times 10^{-6}$
& $(1.9^{+0.5+0.1+0.1}_{-0.3-0.1-0.1})\times 10^{-7}$ & $(8.6^{+2.4+0.2+0.6}_{-2.1-0.7-0.5})\times 10^{-5}$\\
\hline\hline
\end{tabular}
\end{table}
\begin{figure}[!htbh]
\begin{center}
\setlength{\abovecaptionskip}{0pt}
\centerline{
\hspace{-1cm}\subfigure{\epsfxsize=9cm \epsffile{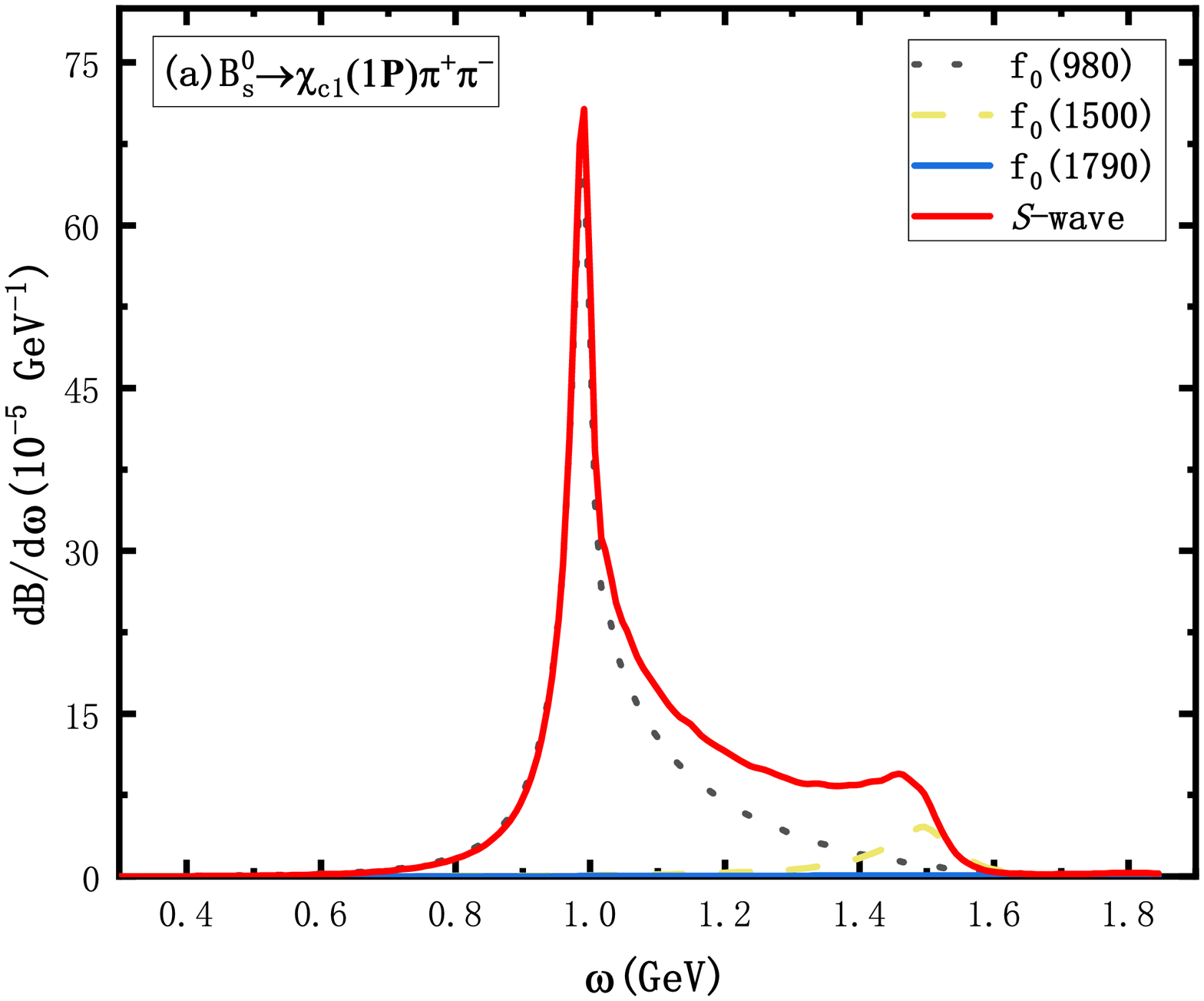} }
\hspace{-1cm}\subfigure{ \epsfxsize=9cm \epsffile{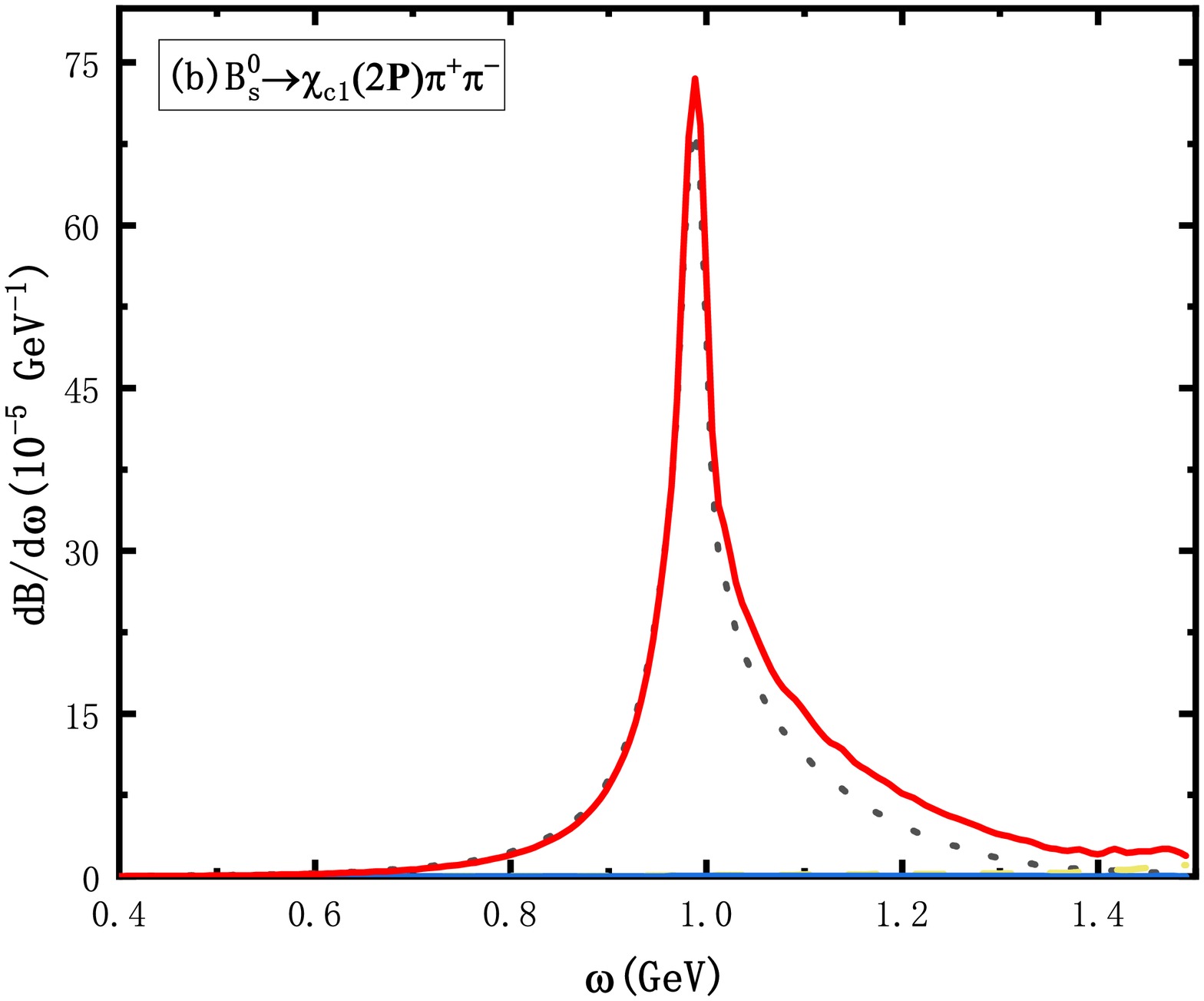}}}
\vspace{0.01cm}
\caption{The $\omega$ dependence of the differential decay branching ratios $dB/d\omega$ for the decay modes (a) $B^0_s\rightarrow \chi_{c1}(1P) \pi^+\pi^-$
and (b) $B^0_s\rightarrow \chi_{c1}(2P) \pi^+\pi^-$.
 The contributions from $f_0(980)$,  $f_0(1500)$ and $f_0(1790)$ components are shown by the dotted gray, dashed khaki and  solid blue curves, respectively,  while the solid red curves represent the total $S$-wave contributions.}
 \label{fig:BSPIPI}
\end{center}
\end{figure}

As mentioned in the previous section,
three resonances were considered in the $B_s^0\rightarrow \chi_{c1}\pi^+\pi^-$ decay,
namely, $f_0(980)$, $f_0(1500)$, and $f_0(1790)$.
The calculated branching ratios of concerned resonances are collected in Table~\ref{tab:brpipi}.
The last column corresponds to the  total $S$-wave branching ratios.
It is evident that the largest contribution comes from the $f_0(980)$ component,
which accounts for $69\% \ (83\%)$ of the $\chi_{c1}(1P(2P))$ mode.
The contributions from high resonances suffer serious suppression since the pole masses
are approaching the upper bound of the two pion invariant mass spectra.
In particular, the  nominal mass of $f_0(1790)$
falls outside the kinematically allowed mass range for the $\chi_{c1}(2P)$ mode
and the residual contribution in the tail region  of the BW function,
also known as virtual contribution, is expected to be fairly small.
The corresponding  branching ratio is predicted to be of order $10^{-7}$,
 much smaller than that of $B_{s}\rightarrow \chi_{c1} f_0(1790)(\rightarrow \pi^+\pi^-)$.
The total $S$-wave branching ratios for the two channels can reach the  $10^{-4}$ level,
to be compared with those of $J/\psi$ modes~\cite{prd91094024,epjc77199},
which is large enough to permit a measurement.

In Fig.~\ref{fig:BSPIPI}, we track the differential branching ratios for various resonances
 as a function of the $\pi^+\pi^-$ invariant mass,
 which we vary from $2m_{\pi}$  up to $M-m$.
The dotted gray, dashed khaki and solid blue curves correspond to the $f_0(980)$, $f_0(1500)$, and $f_0(1790)$ resonances, respectively,
while the solid red curves represent the total $S$-wave contributions.
Note that the mass difference between $\chi_{c1}(1P)$ and $\chi_{c1}(2P)$  causes significant differences in the range
spanned in the respective decay modes.
One can see a clear signal from the $f_0(980)$ resonance, accompanied by $f_0(1500)$, 
while the amount of $f_0(1790)$ is less than $1\%$ of the total $S$-wave contributions.
A dip in Fig.~\ref{fig:BSPIPI}(a) in the invariant mass region of 1.2--1.4 GeV is ascribed to
 the  interference between the  $f_0(980)$  and $f_0(1500)$  channels.
 However, such a dip is not observed in Fig.~\ref{fig:BSPIPI}(b)
 because the $f_0(1500)$ component suffers strong phase space suppression for the $\chi_{c1}(2P)$ mode.
\subsection{ $B^0\rightarrow \chi_{c1}\pi^+\pi^-$}
\begin{table}
\caption{ Branching ratios  of $S$-wave resonant contributions to  the $B^0\rightarrow \chi_{c1}(1P,2P)f_{0}(500) ( \rightarrow\pi\pi)$ decays.
The theoretical errors correspond to the uncertainties due to the shape parameters $\omega_B$, the hard scale $t$, and  the Gegenbauer moment $a_{\pi\pi}$, respectively. }
\label{tab:brpipi500}
\begin{tabular}[t]{lcc}
\hline\hline
Modes &$\mathcal{B}(\text{BW})$&$ \mathcal{B}(\text{Bugg} )$ \\ \hline 
$B^0\rightarrow \chi_{c1}(1P) f_{0}(500) (\rightarrow\pi^+\pi^- )$  & $(2.9^{+1.0+0.1+1.1}_{-0.7-0.1-0.5})\times 10^{-6}$ & $(2.8^{+0.6+0.0+0.4}_{-0.6-0.0-0.3})\times 10^{-6}$ \\  
$B^0\rightarrow \chi_{c1}(2P) f_{0}(500) (\rightarrow\pi^+\pi^- )$  & $(2.8^{+0.9+0.0+0.5}_{-0.7-0.1-0.0})\times 10^{-6}$ &  $(1.7^{+0.4+0.0+0.1}_{-0.3-0.0-0.0})\times 10^{-6}$ \\ 
\hline\hline
\end{tabular}
\end{table}
\begin{figure}[!htbh]
\begin{center}
\setlength{\abovecaptionskip}{0pt}
\centerline{
\hspace{-1cm}\subfigure{\epsfxsize=9cm \epsffile{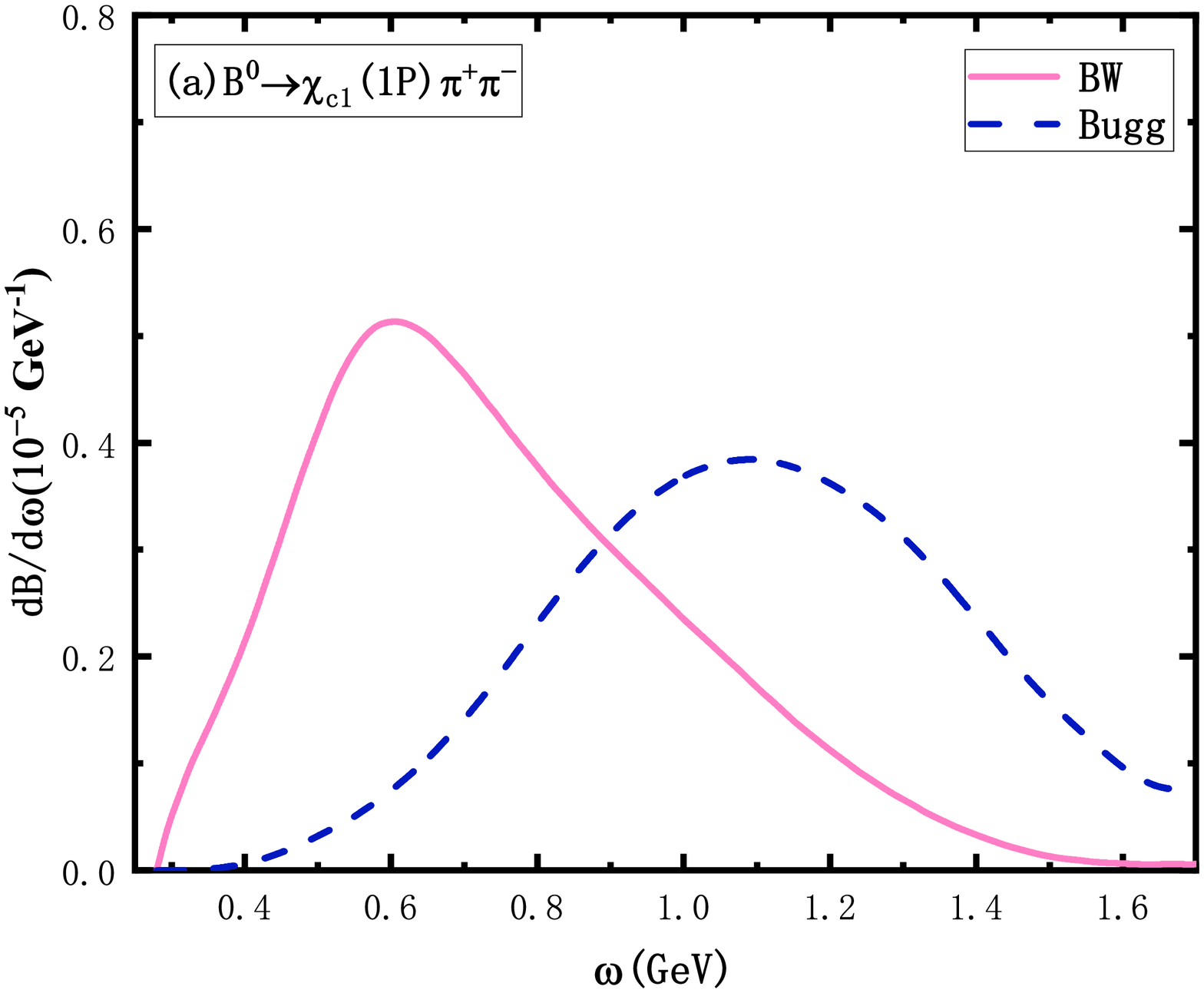} }
\hspace{-1cm}\subfigure{ \epsfxsize=9cm \epsffile{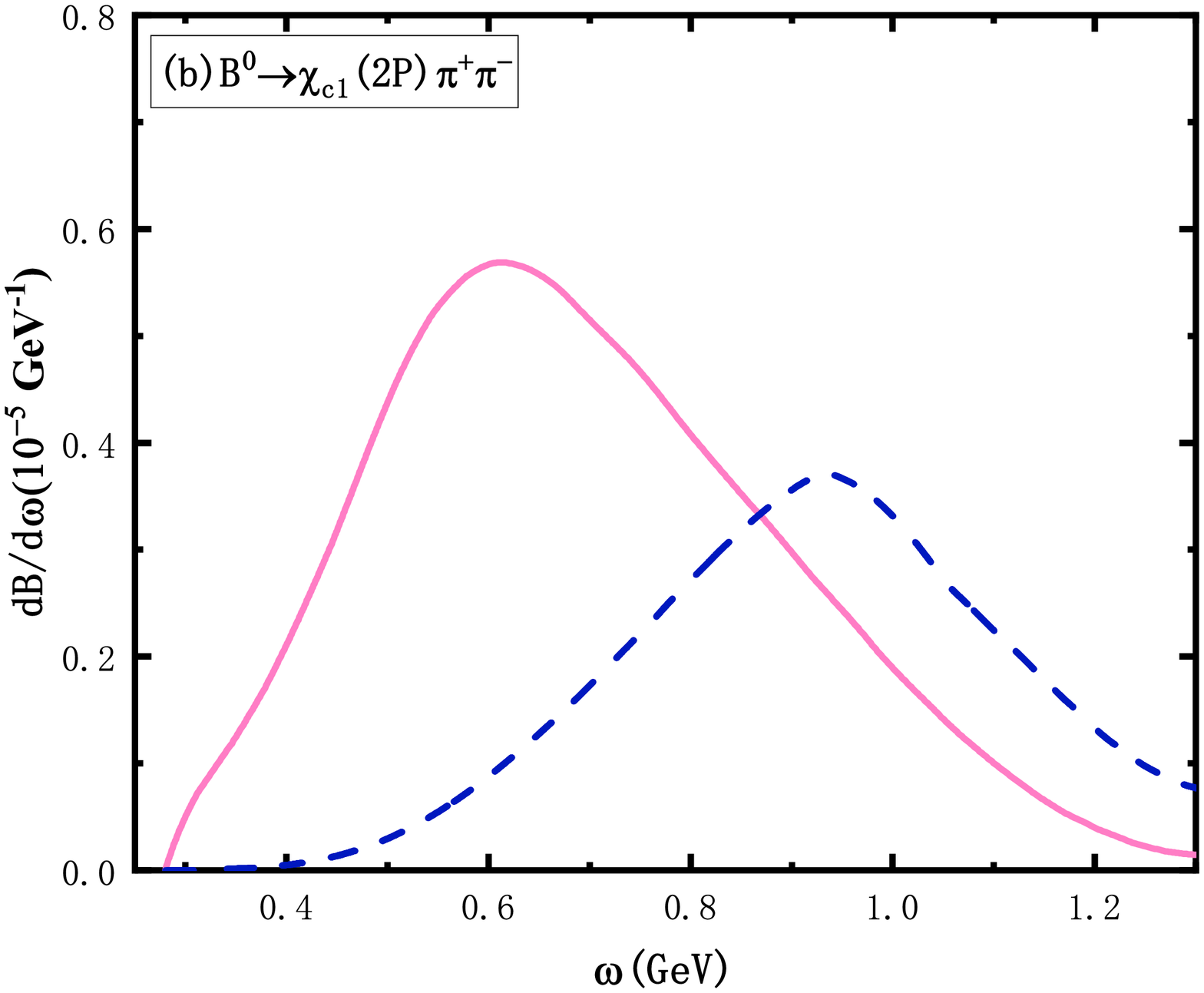}}}
\vspace{0.01cm}
\caption{ (a) d$\mathcal{B}/d \omega$ invariant-mass distribution for the  $B^0\rightarrow \chi_{c1}(1P)f_{0}(500)(\rightarrow\pi^+\pi^-)$ decay
with two different descriptions for the $f_0(500)$ resonance. Similar curves are shown in (b) but for the $B^0\rightarrow \chi_{c1}(2P)f_{0}(500)(\rightarrow\pi^+\pi^-)$ mode.
 The BW and Bugg models are shown by  the solid pink and dashed blue lines, respectively.}
 \label{fig:BPIPIpp}
\end{center}
\end{figure}

The branching ratios of $B^0\rightarrow \chi_{c1}(1P,2P)f_0(500)(\rightarrow \pi^+\pi^-)$ decays,
 calculated for both  BW and  Bugg models, are presented for comparison in Table~\ref{tab:brpipi500}.
 The dependence of differential branching ratios as a function of the invariant mass is also
shown in Fig.~\ref{fig:BPIPIpp} for the aforementioned two line shapes,
where the solid pink and dashed blue lines correspond to the BW and Bugg models, respectively.
The two curves have a broad bump with different behaviors in shape.
In the BW model,
the peak at the resonance is usually highly dominating and
the majority of the resonant contribution  focuses naturally on the mass range of $f_0(500)$.
Although $f_0(500)$ is very broad,  the predominated  mass region still far away from the the upper bound of allowed phase space, $M-m$.
However, the strength of the Bugg model is slightly small, as seen in Fig.\ref{fig:BPIPIpp},
the spectrum bump is located  above the pole mass of $f_0(500)$
 because the substantial coupling of $f_0(500)$ to $KK$ and $\eta\eta$ are included.
As a consequence, the contribution from the high-mass regions can not be ignored. 
From Table~\ref{tab:brpipi500}, we find   $\mathcal{B}(B^0\rightarrow \chi_{c1}(1P)f_0(500)(\rightarrow \pi^+\pi^-))$ and
 $\mathcal{B}(B^0\rightarrow \chi_{c1}(2P)f_0(500)(\rightarrow \pi^+\pi^-))$ are  of comparable size for the BW model,
but the latter is relatively small owing to the phase space suppression for the Bugg one.

\subsection{ $B_{(s)}^0\rightarrow \chi_{c1}K\pi$}
\begin{table}
\caption{ Branching ratios of various $S$-wave components to  the $B_{(s)}\rightarrow \chi_{c1}(1P,2P) K\pi$ decays.
The theoretical errors correspond to the uncertainties due to the shape parameters $\omega_{B_{(s)}}$, the hard scale $t$, and  the Gegenbauer moments $B_{1,3}$, respectively. }
\label{tab:brkpi}
\begin{tabular}[t]{lccc}
\hline\hline
Modes & $\mathcal{B}(R=K_{0}^{\ast}(1430))$& $\mathcal{B}(\text{LASS NR})$& $\mathcal{B}(\text{S-wave})$\\ \hline
$B^0_s\rightarrow \chi_{c1}(1P) K^-\pi^+ $  & $(3.8^{+0.7+0.3+0.5}_{-0.6-0.2-0.4})\times 10^{-6}$ & $(3.4^{+0.9+0.3+0.4}_{-0.7-0.2-0.4})\times 10^{-6}$  & $(6.9^{+1.9+0.5+0.9}_{-1.6-0.5-0.8})\times 10^{-6}$  \\
$B^0_s\rightarrow \chi_{c1}(2P) K^-\pi^+$  & $(1.5^{+0.2+0.0+0.1}_{-0.3-0.1-0.2})\times 10^{-6}$ & $(1.8^{+0.6+0.1+0.4}_{-0.3-0.1-0.1})\times 10^{-6}$ & $(4.4^{+1.2+0.2+0.6}_{-0.9-0.3-0.5})\times 10^{-6}$\\
$B^0\rightarrow \chi_{c1}(1P) K^+\pi^- $  & $(5.1^{+0.3+0.1+0.5}_{-0.6-0.2-0.6})\times 10^{-5}$ & $(5.3^{+1.0+0.1+0.5}_{-1.0-0.2-0.5})\times 10^{-5}$& $(1.1^{+0.2+0.1+0.2}_{-0.2-0.0-0.1})\times 10^{-4}$\\
$B^0\rightarrow \chi_{c1}(2P) K^+\pi^- $  & $(1.6^{+0.2+0.0+0.1}_{-0.3-0.1-0.2})\times 10^{-5}$ & $(2.8^{+0.5+0.0+0.4}_{-0.6-0.1-0.4})\times 10^{-5}$& $(6.4^{+1.2+0.0+0.7}_{-1.2-0.1-0.7})\times 10^{-5}$ \\
\hline\hline
\end{tabular}
\end{table}
\begin{figure}[!htbh]
\begin{center}
\setlength{\abovecaptionskip}{0pt}
\centerline{
\hspace{-1cm}\subfigure{\epsfxsize=9cm \epsffile{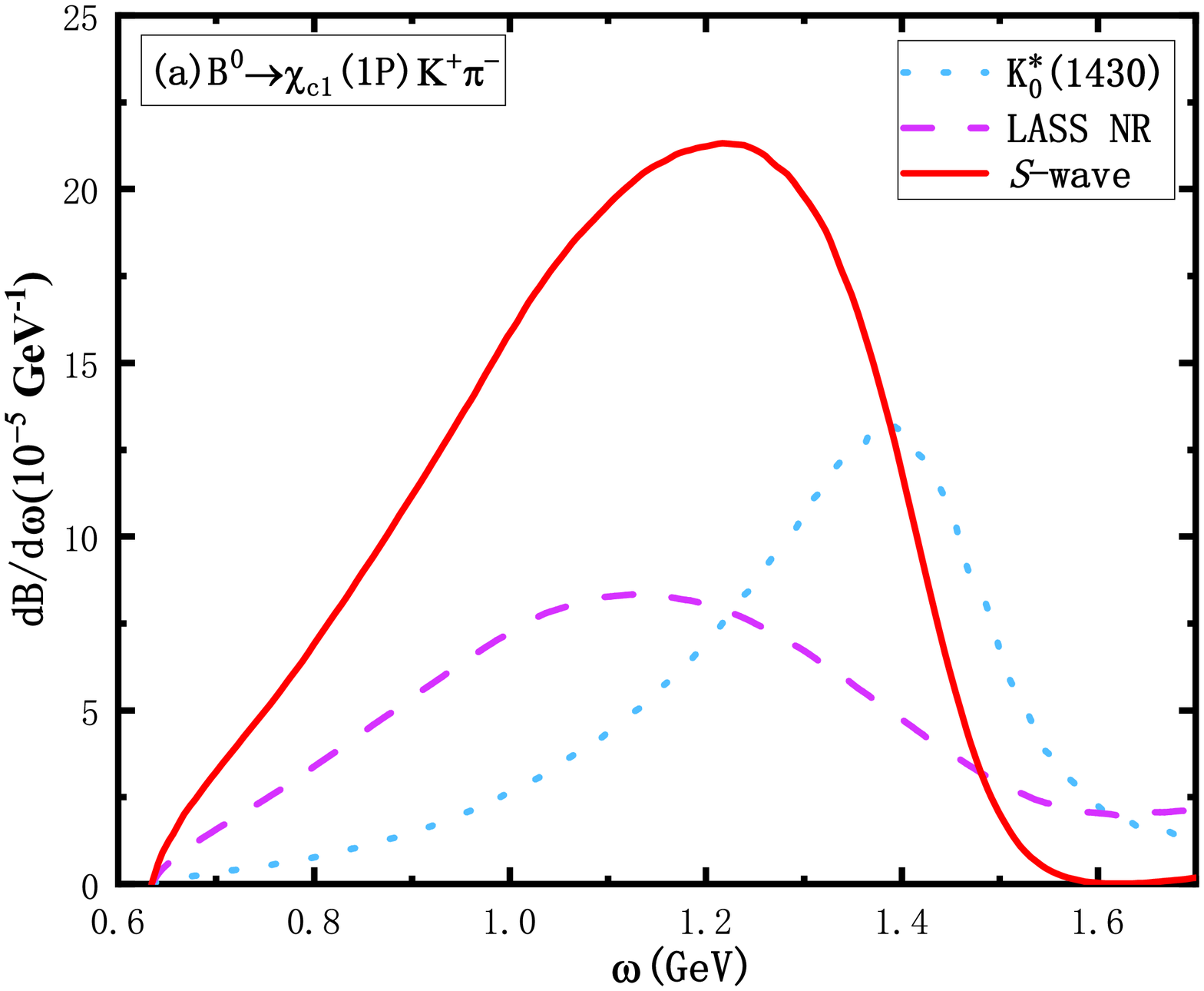} }
\hspace{-1cm}\subfigure{ \epsfxsize=9cm \epsffile{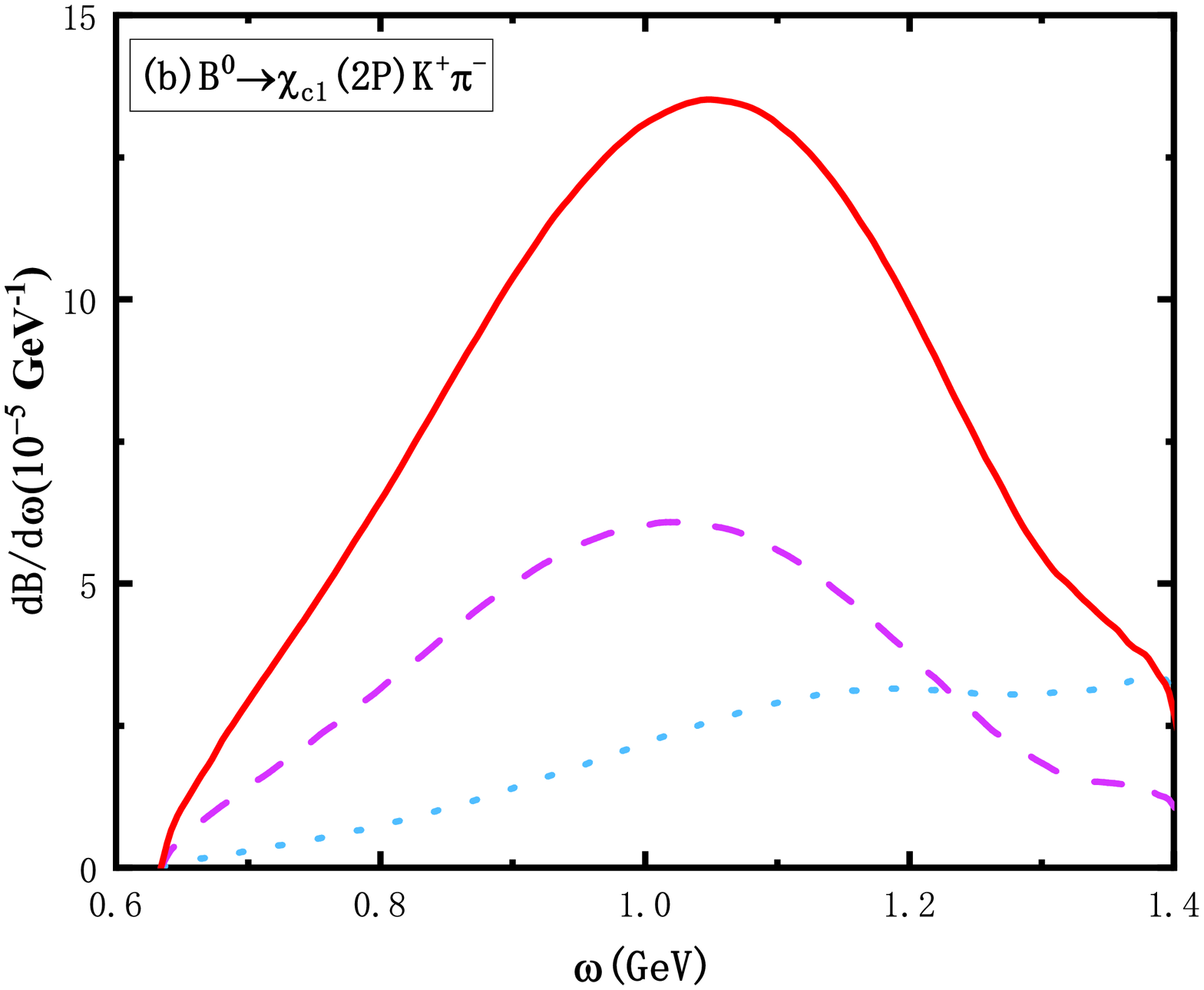}}}
\vspace{0.01cm}
\centerline{
\hspace{-1cm}\subfigure{\epsfxsize=9cm \epsffile{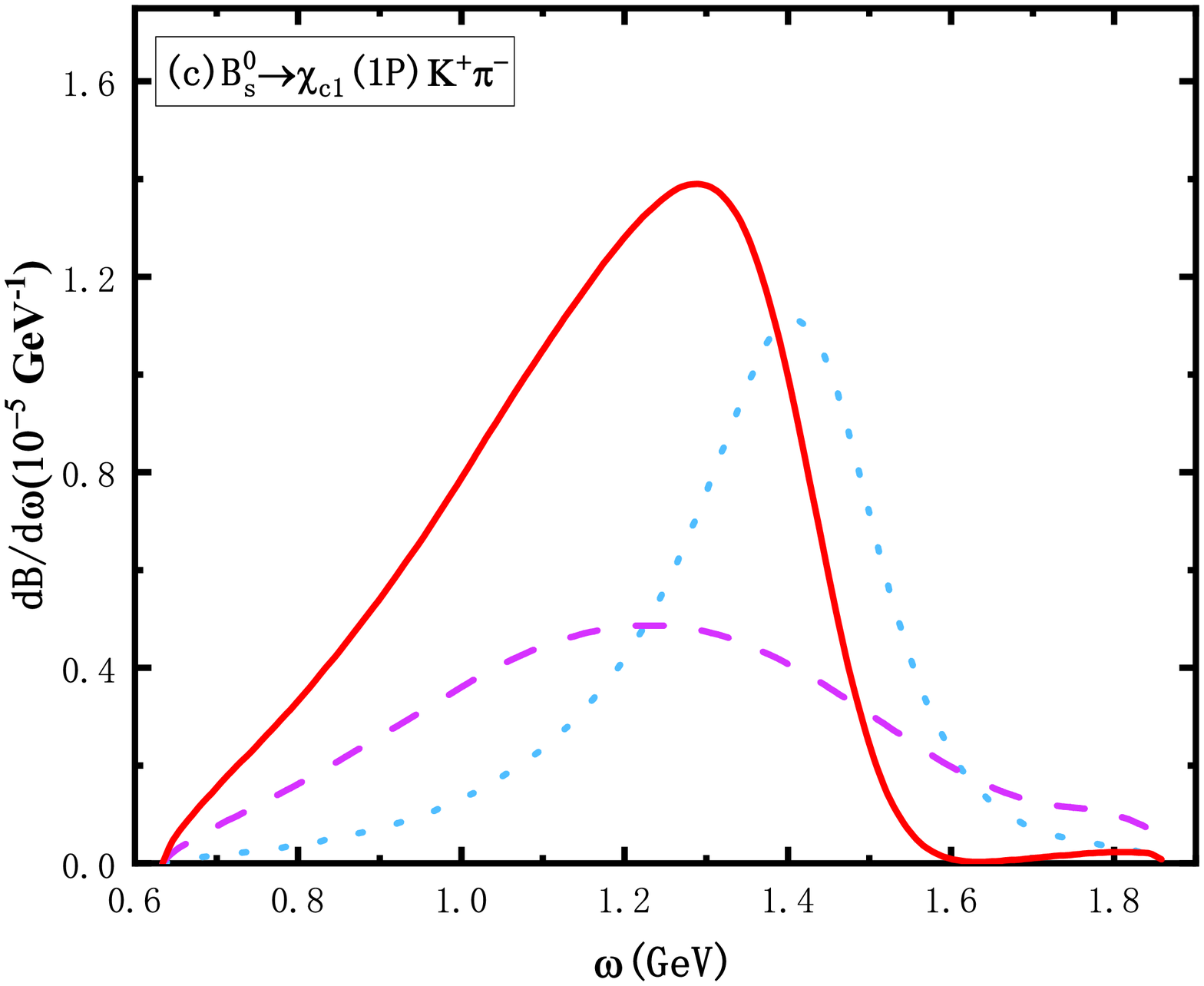} }
\hspace{-1cm}\subfigure{ \epsfxsize=9cm \epsffile{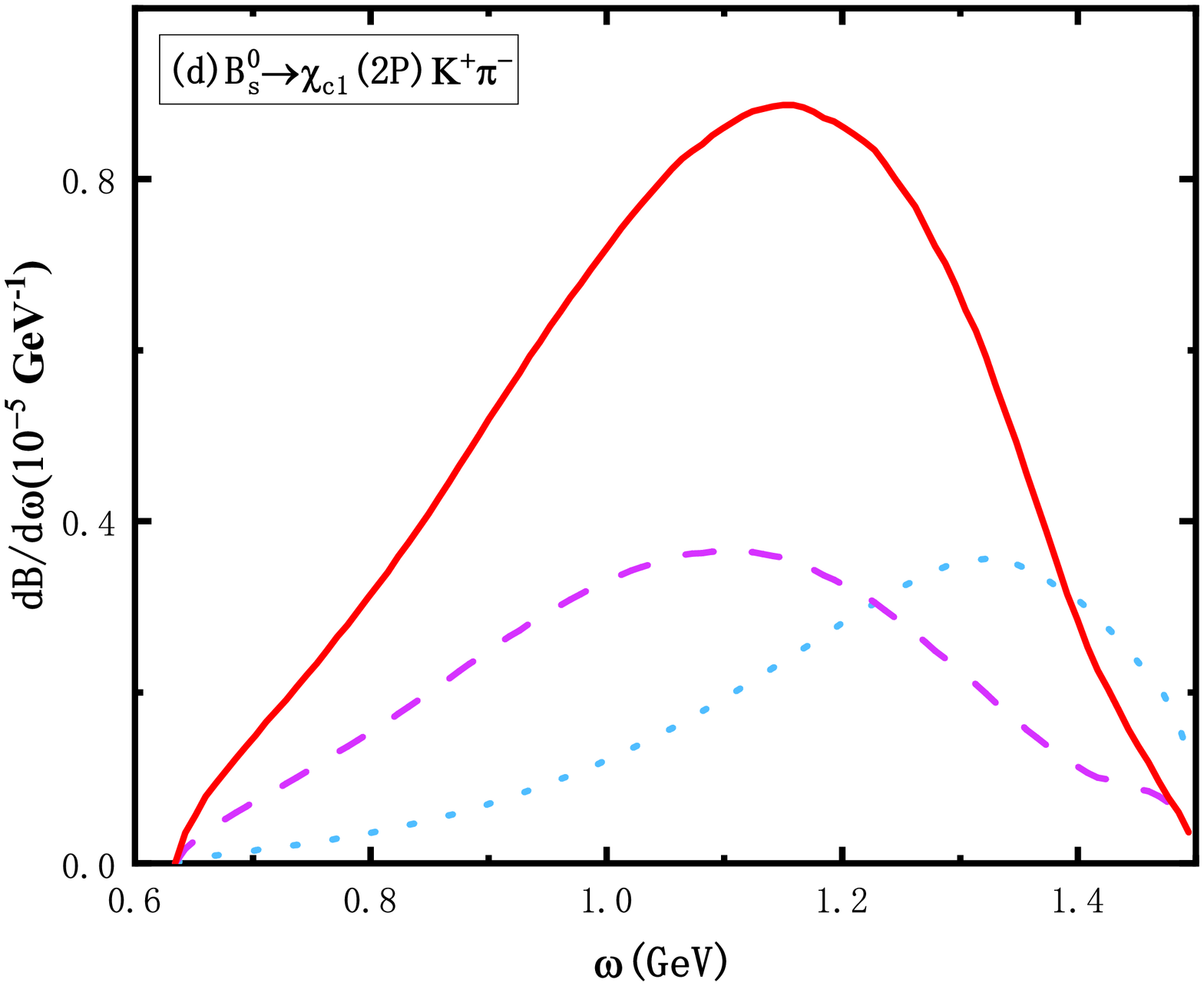}}}
\vspace{0.01cm}
\caption{$S$-wave  contributions to the differential branching ratios of the modes (a) $B^0\rightarrow \chi_{c1}(1P) K^+\pi^- $,
(b) $B^0\rightarrow \chi_{c1}(2P) K^+\pi^- $, (c) $B^0_s\rightarrow \chi_{c1}(1P) K^-\pi^+ $, and (d) $B^0_s\rightarrow \chi_{c1}(2P) K^-\pi^+ $.
 The dotted blue, dashed violet, and solid red show the contributions from resonances $K_{0}^{\ast}(1430)$, LASS NR and their combinatorial, respectively.}
 \label{fig:BSKPI}
\end{center}
\end{figure}

Since the LASS description for the $K\pi$ $S$-wave  contained both resonant and nonresonant components  as shown in Eq.~(\ref{eq:sform}),
we summarize the branching ratios of $K^*(1430)$ resonant and nonresonant component as well as the total $S$-wave contribution
in Table~\ref{tab:brkpi} separately.
The invariant mass dependence of the differential decay rates for the two components is shown in Fig.~\ref{fig:BSKPI}.
Analogous to the pattern of the $B_{(s)}\rightarrow J/\psi K\pi$ decays in our previous analysis~\cite{prd97033006},
the contributions from the resonant and nonresonant components in this work are  comparable in size.
Hence, the nonresonant contributions also play an essential role in $B_{(s)}\rightarrow \chi_{c1}K\pi$ decays.
The constructive interference between the  resonant and nonresonant contributions lead  to
large $S$-wave  branching ratios as shown in Fig.~\ref{fig:BSKPI},
especially  the  $B^0\rightarrow \chi_{c1}(1P)K^+\pi^-$ mode has a large branching ratio of order  $10^{-4}$.
The corresponding $B_s$ channels have relatively small branching ratios $(10^{-6})$ comparing with the $B^0$ modes
due to the CKM suppression $|V_{cd}/V_{cs}|\sim \lambda$.

From the  fit fraction of the $K^*_0(1430)$ component in  $\bar{B}^0\rightarrow \chi_{c1}K^-\pi^+$ decay analyzed in the isobar
model~\cite{prd78072004}  and the three-body branching ratio $\mathcal{B}(\bar{B}^0\rightarrow \chi_{c1}K^-\pi^+)=(3.83\pm0.10\pm0.39)\times 10^{-4}$ measured by Belle~\cite{prd78072004}, we obtain
\begin{eqnarray}
\mathcal{B}(B^0\rightarrow \chi_{c1}K^*_0(1430)(\rightarrow K^+\pi^-))_{\text{expt}}&=&\left\{
\begin{aligned}
&(8.6\pm 2.4)\times 10^{-5} \quad\quad\quad  &\text{S1}, \nonumber\\ 
&(7.1\pm 2.1)\times 10^{-5} \quad\quad\quad  &\text{S2},  \nonumber\\ 
\end{aligned}\right.
\end{eqnarray}
where S1 and S2 denote the two solutions from
 single- and double-$Z^+$ resonance scenarios in the $\chi_{c1}\pi^+$ invariant mass distribution, respectively.
One can see from Table~\ref{tab:brkpi} that the predicted branching ratio
$\mathcal{B}(\bar{B}^0\rightarrow \chi_{c1}K^*_{0}(1430)(\rightarrow K^-\pi^+))=(5.1^{+0.6}_{-0.8})\times 10^{-5}$
is  compatible with the above two solutions  within errors.

The $BABAR$ collaboration  measured the three-body branching ratio,
$\mathcal{B}(\bar{B}^0\rightarrow \chi_{c1}K^-\pi^+)=(5.11\pm0.14\pm0.28)\times 10^{-4}$~\cite{Lees:2011ik}.
Meanwhile, the $S$-, $P$-, and $D$-wave fractions are fitted from the analysis of the  $K\pi$ mass spectra.
It is found that the $S$-wave fraction in $\bar{B}^0\rightarrow \chi_{c1} K^-\pi^+$ is larger than
the  corresponding $J/\psi$ and $\psi(2S)$ modes.
 Multiplying the three-body branching ratio quoted above  by the $S$-wave fraction  $f_{S}=(40.4\pm2.2)\%$,
 we obtained the $S$-wave branching ratio
 \begin{eqnarray}
\mathcal{B}(\bar{B}^0\rightarrow \chi_{c1} (K^-\pi^+)_{\text{S}})= \mathcal{B}(\bar{B}^0\rightarrow \chi_{c1} K^-\pi^+)\times f_S=2.1^{+0.1}_{-0.2}\times 10^{-4},
 \end{eqnarray}
which is twice our prediction in Table~\ref{tab:brkpi}.
We note that the three-body branching ratio measured by $BABAR$ is typically
larger than the previous measurement by Belle~\cite{prd78072004} and also larger than the updated measurement of
$\mathcal{B}(\bar{B}^0\rightarrow \chi_{c1}K^-\pi^+)=(4.97\pm0.12\pm0.28)\times 10^{-4}$ in~\cite{prd93052016}.
As an aside, the quoted uncertainty for the fitted $S$-wave fraction is statistical only,
thereby improving  precision on both the theoretical and experimental values would be highly desirable.

\subsection{ $B_s^0\rightarrow \chi_{c1}K^+K^-$}
\begin{table}
\caption{ Branching ratios  of $S$-wave resonant contributions to  the $B^0_{s}\rightarrow \chi_{c1}(1P,2P) K^+K^-$ decays.
The theoretical errors correspond to the uncertainties due to the shape parameter $\omega_{B_s}$, the hard scale $t$, and  the Gegenbauer moment $a_{KK}$, respectively.}
\label{tab:brkk}
\begin{tabular}[t]{lcccc}
\hline\hline
Modes & $\mathcal{B}(R=f_{0}(980))$ & $\mathcal{B}(R=f_{0}(1370))$ & $\mathcal{B}(\text{S-wave})$\\ \hline
$B^0_s\rightarrow \chi_{c1} (1P)K^+K^- $ & $(3.5^{+1.1+0.5+0.3}_{-0.8-0.3-0.2})\times 10^{-5}$& $(9.2^{+0.7+0.8+0.3}_{-1.1-0.7-0.3})\times 10^{-6}$
& $(3.9^{+1.0+0.5+0.2}_{-0.8-0.3-0.2})\times 10^{-5}$ \\
$B^0_s\rightarrow \chi_{c1}(2P) K^+K^- $ & $(2.7^{+0.7+0.2+0.2}_{-0.6-0.3-0.2})\times 10^{-5}$& $(2.4^{+0.5+0.0+0.1}_{-0.5-0.2-0.2})\times 10^{-6}$
& $(2.4^{+0.7+0.2+0.2}_{-0.6-0.2-0.2})\times 10^{-5}$ \\
\hline\hline
\end{tabular}
\end{table}
\begin{figure}[!htbh]
\begin{center}
\setlength{\abovecaptionskip}{0pt}
\centerline{
\hspace{-1cm}\subfigure{\epsfxsize=9cm \epsffile{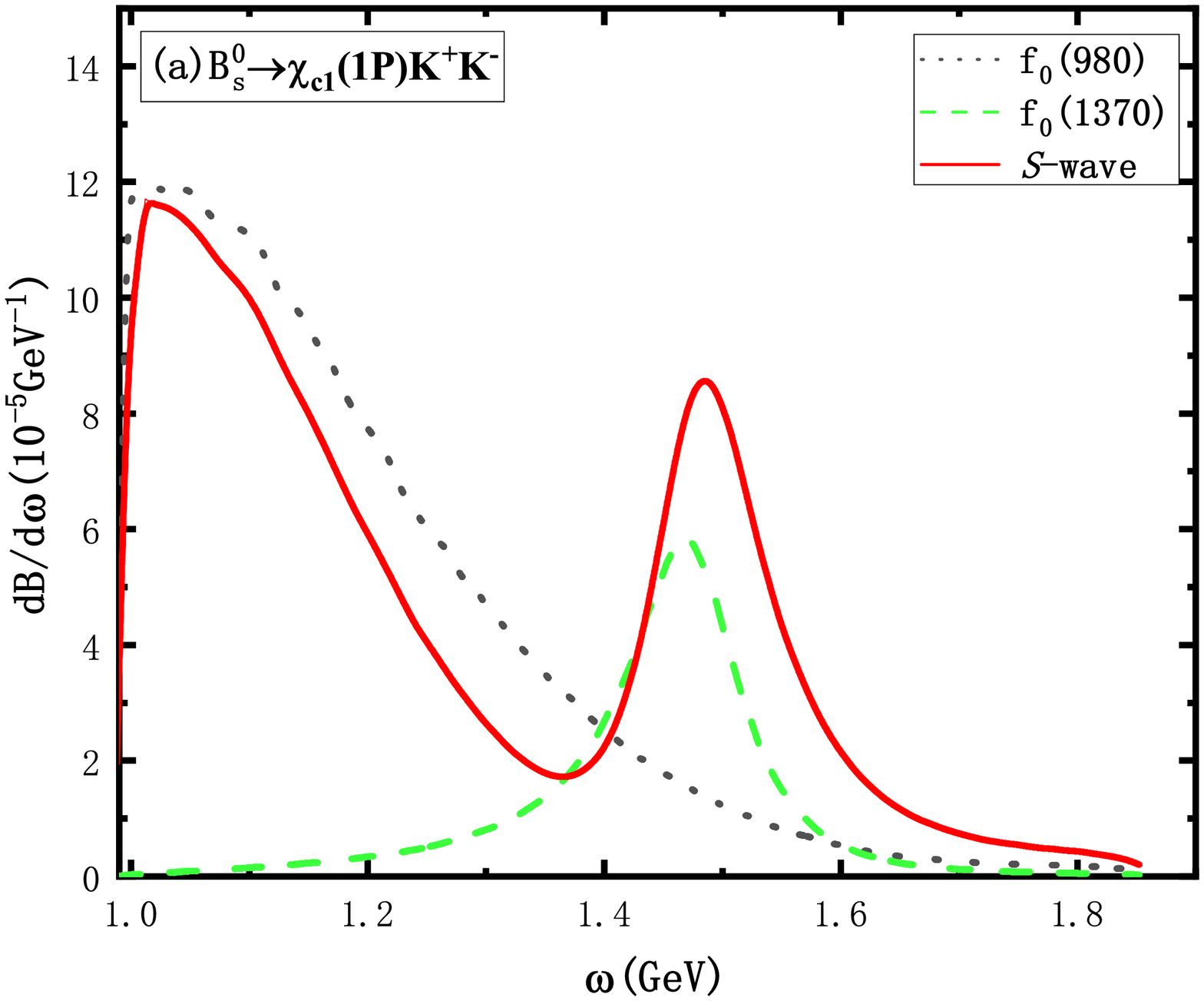} }
\hspace{-1cm}\subfigure{ \epsfxsize=9cm \epsffile{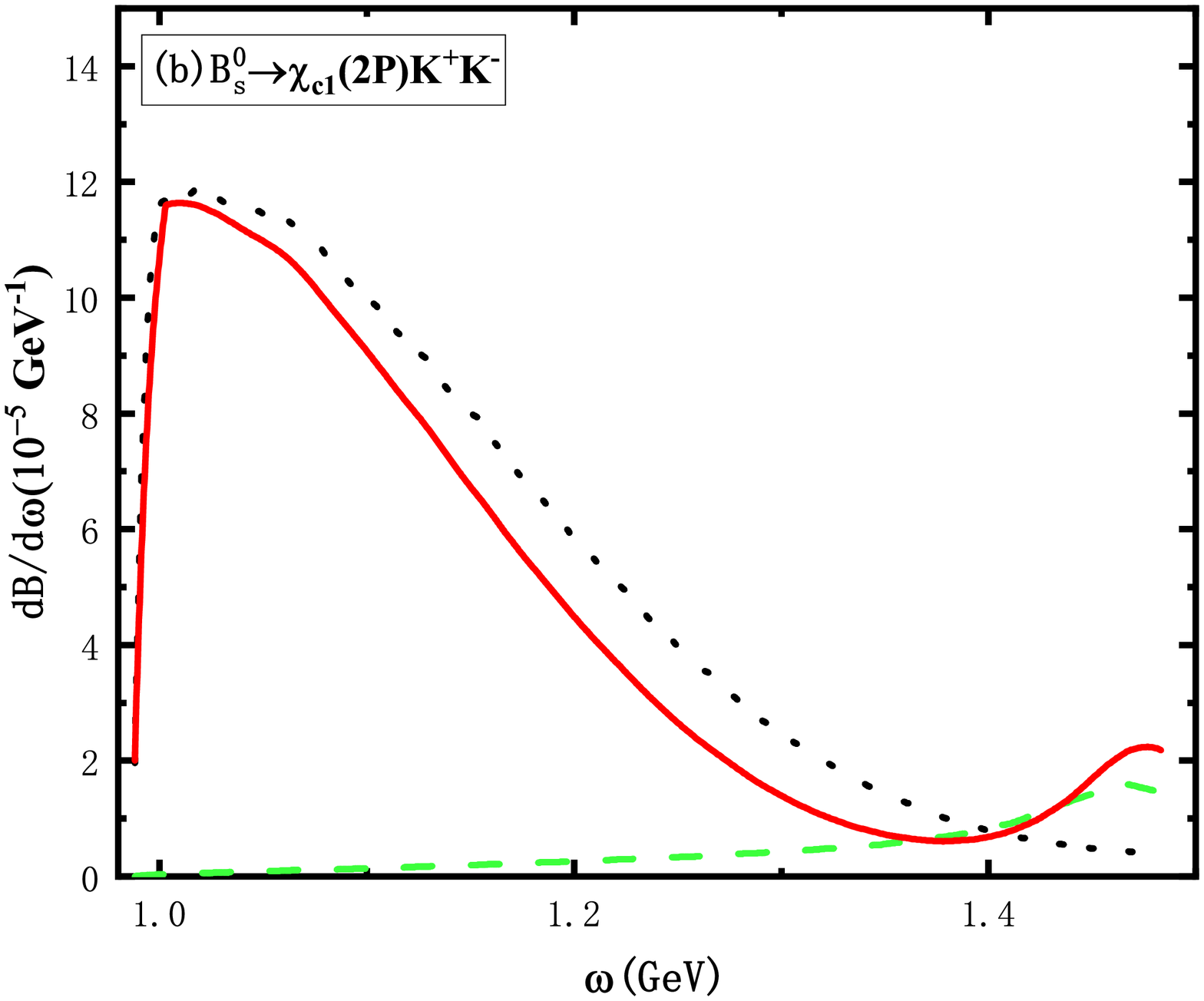}}}
\vspace{0.01cm}
\caption{Various resonance contributions to the differential branching ratios of the modes
(a) $B^0_{s}\rightarrow \chi_{c1}(1P) K^+K^-$  and (b) $B^0_{s}\rightarrow \chi_{c1}(2P) K^+K^-$.
The dotted gray, dashed green and solid red curves show the resonances $f_0(980)$, $f_0(1370)$,
and their combinatorial contributions, respectively.}
 \label{fig:BSKK}
\end{center}
\end{figure}

We next turn to the $B^0_{s}\rightarrow \chi_{c1}K^+K^-$ decay which receives two resonant contributions from
$f_0(980)$ and $f_0(1370)$ in the $K^+K^-$ invariant mass spectrum.
The  predicted branching ratios are depicted in Table~\ref{tab:brkk},
while the corresponding differential   distributions over $\omega$ are plotted in Fig.~\ref{fig:BSKK}.
The red (solid) curves denote the total contribution,
while individual terms are given by the gray (dotted) lines for $f_0(980)$ and green (dashed) lines for $f_0(1370)$.
It is clear that the $S$-wave contributions are dominated by $f_0(980)$,  while the $f_0(1370)$ component is  several times smaller.
The total $S$-wave branching ratios  reach the order of $10^{-5}$, which are comparable with those of $\pi\pi$ modes in Table~\ref{tab:brpipi}.
Since $f_0(980)$ can decay into a kaon  or a pion pair, we can estimate the  relative branching ratios:
\begin{eqnarray}\label{eq:r}
\mathcal{R}=\frac{\mathcal{B}(B^0_s\rightarrow \chi_{c1}f_0(980)(\rightarrow K^+K^-))}
{\mathcal{B}(B^0_s\rightarrow \chi_{c1}f_0(980)(\rightarrow \pi^+\pi^-))}.
\end{eqnarray}
Combining Tables~\ref{tab:brpipi} and~\ref{tab:brkk}, the ratio $\mathcal{R}$ is predicted to be
$0.46^{+0.24}_{-0.19}(0.37^{+0.18}_{-0.14})$ for the $1P(2P)$ state mode,
which is comparable with  our previous prediction for that of $J/\psi$ with $\mathcal{R}(J/\psi)=0.37^{+0.23}_{-0.13}$~\cite{epjc79792}.
In the narrow-width limit, Eq.(\ref{eq:r}) simplifies to
\begin{eqnarray}
\mathcal{R}\approx\frac{\mathcal{B}(f_0(980)\rightarrow K^+K^-)}{\mathcal{B}(f_0(980)\rightarrow \pi^+\pi^-)},
\end{eqnarray}
where the common term $\mathcal{B}(B^0_s\rightarrow \chi_{c1}f_0(980))$ in the numerator and denominator  cancel out.
The weighted average of $\mathcal{R}$ from $BABAR$~\cite{Aubert:2006nu} and BES~\cite{Ablikim:2005kp} measurements yields $\mathcal{R}_{\text{expt}}=0.35^{+0.15}_{-0.14}$~\cite{prd90012003}. 
Our estimates turn out to be consistent with this average value.

From Tables~\ref{tab:brpipi}--\ref{tab:brkk}
one can see that  the branching ratios of the lower mass resonances between $\chi_{c1}(1P)$ and  $\chi_{c1}(2P)$ modes are comparable,
whereas in the case of  the higher mass resonances, the corresponding branching ratios for the latter are typically smaller 
due to the phase space suppression.
Since the $\chi_{c1}(2P)$ modes received less theoretical and experimental attention, we wait for future comparison,
which may help us to further clarify the structure of $X(3872)$.
\section{ conclusion}\label{sec:sum}
We have analyzed the three-body decays $B_{(s)}\rightarrow \chi_{c1} hh'$ in the $hh'$ invariant mass spectrum with $S$-wave configuration in the perturbative QCD  approach.
The final-state pseudoscalar meson $h^{(')}$ is restricted to be a kaon or a pion.
The $S$-wave contributions are parametrized into the  timelike form factors involved in the two meson DAs,
which have been well established in the corresponding $J/\psi$ decays.

The strange scalar form factor for the $\pi\pi$ pair is described by the coherent sum of three
scalar resonances $f_0(980)$, $f_0(1500)$, and $f_0(1790)$.
Except for $f_0(980)$, parametrized by a Flatt\'{e} line shape, the latter two resonances are modeled by the Breit-Wigner function.
The nonstrange scalar form factor contains only the $f_0(500)$  resonance,
which is modeled with two alternative shapes, the BW and Bugg formulas.
 Although the resultant invariant mass distributions for the two models show a different behavior,
 the integrated branching ratios  over the entire phase space are comparable.
The $K\pi$ timelike form factor is described by the conventional LASS parametrization,
which consists of $K^*_{0}(1430)$ resonance together with an effective range nonresonant component.
It is found that the contributions from the two pieces are of comparable size.
In the $KK$  sector, the corresponding  timelike form factor is parametrized by a linear combination of the
$f_0(980)$ and $f_0(1370)$ resonances, where the latter  is also modeled by the BW line shape.

By using the well established  two-meson DAs,
we have calculated the branching ratios together with the differential distributions of various components
 in the processes under consideration.
 The branching ratio of $B\rightarrow \chi_{c1}K^*_{0}(1430)(\rightarrow K^+\pi^-)$
 is predicted to be $(5.1^{+0.6}_{-0.8})\times 10^{-5}$, which is in agreement with the Belle measurement,
 while the obtained $S$-wave branching ratio is smaller than the $BABAR$ data by a factor of 2.
 The branching ratios  for some Cabibbo-favored decays  are large  of order $10^{-4}$,
 whereas those of the Cabibbo-suppressed  ones are at least lower by an order of
magnitude because of the smaller CKM matrix elements.
The obtained distribution for the various components contributions to the
 considered decays can  be tested by future experimental measurements.
Additionally, our predictions on the $\chi_{c1}(2P)$ modes
 could help to understand the $X(3872)$ properties.

\section*{ACKNOWLEDGMENTS}
This work is supported by National Natural Science Foundation
of China under Grants No. 12075086 and  No. 11605060 and the Natural Science Foundation of Hebei Province
under Grants No.A2021209002 and  No.A2019209449.

\begin{appendix}
\section{DETAILS FOR DERIVING $\chi_{c1}(2P)$ CHARMONIUM DAS}
We begin with  the momentum-space radial wave function which can be written as a Fourier transform of the position-space expression $\Psi_{nlm}(\vec{r})$,
\begin{eqnarray} \label{eq:kk}
\Psi(k)=\int_{-\infty}^{\infty}\Psi_{nlm}(\vec{r})e^{-\vec{k}
\cdot\vec{r}}d\vec{r},
\end{eqnarray}
where $n$, $l$ , and $m$ stand for main, orbital, and magnetic quantum numbers, respectively.
In the spherical coordinates $(r,\theta,\varphi)$,
the two terms $\Psi_{nlm}(\vec{r})$ and $e^{-\vec{k}\cdot\vec{r}}$ in Eq.~(\ref{eq:kk})  can be written as
\begin{eqnarray}\label{eq:kk1}
\Psi_{nlm}(\vec{r})&=&R_{nl}(r)Y_{lm}(\theta,\varphi), \nonumber\\
e^{-\vec{k}\cdot\vec{r}}&=&e^{-ikr\cos\theta}=
\sum\limits_{l^{'}=0}^\infty\sqrt{4\pi(2l^{'}+1)}(-i)^{l^{'}}j_{l^{'}}(kr)
Y_{l^{'}0}(\theta,0),
\end{eqnarray}
where $j_{l^{'}}(kr)$ is the spherical Bessel function.
Substituting the results of Eq.~(\ref{eq:kk1}) in Eq.~(\ref{eq:kk}),  we obtain
\begin{eqnarray}\label{jj}
\Psi(k)=\sqrt{4\pi(2l^{'}+1)}(-i)^{l}\int_{0}^{\infty}j_{l}(kr)R_{nl}r^{2}dr,
\end{eqnarray}
where the orthogonality property $\int_{0}^{\pi}\int_{0}^{2\pi}Y_{lm}Y_{l^{'}0}\sin\theta d\theta d\varphi=\delta_{ll^{'}}\delta_{m0}$ has been employed.
For the $\chi_{c1}(2P)$ state with quantum numbers $n=3$ and $l=1$,
employing the spherical Bessel function $j_{1}(kr)=\frac{\sin(kr)-kr\cos(kr)}{(kr)^{2}}$
and the radial wave function for a Coulomb potential
$R_{31}(r)\propto r \left(1-\frac{q _{B} r}{6}\right) e^{ -\frac{1}{3} q _{B} r}$ with $q_{B}$ being the Bohr momentum.
The integral of Eq.(\ref{jj}) evaluates to
\begin{eqnarray}\label{mm}
\Psi(k)\propto\frac{  k \left(9 k^2-q _{B}^2\right)}{\left(q _{B}^2+9 k^2\right)^4}.
\end{eqnarray}
Following the similar strategy proposed in Refs.~\cite{plb612215,prd71114008},
we obtain the heavy quarkonium DA which is dependent on the charm quark momentum fraction $x$ after integrating the transverse momentum $k_{T}$,
\begin{eqnarray}
\Phi(x)\sim \int d^{2}k_{T}\Psi(x,k_{T})\propto x(1-x)\times \left\{\frac{\left((1-2 x)^2(1-x) x\right)^{3/2}}{\left(1-\frac{4}{9} \left(v^2-9\right) (x-1) x\right)^3} \right\},
\end{eqnarray}
where $v=q_{B}/m_{c}$ is the charm quark velocity.
Similar to the DAs of $\chi_{c1}(1P)$~\cite{prd97033001},
we can propose that of $\chi_{c1}(2P)$ as $\Psi(x)\propto \Phi_{asy}(x)\mathcal{T}(x)$ with
\begin{eqnarray}\label{t2}
\mathcal{T}(x)=\left\{\frac{\left((1-2 x)^2(1-x) x\right)^{3/2}}{\left(1-\frac{4}{9} \left(v^2-9\right) (x-1) x\right)^3} \right\}^{1-v^{2}},
\end{eqnarray}
where the power $1 -v^{2} $ denotes the small relativistic
corrections to the Coulomb wave functions. In the numerical calculation, we take $v^{2}=0.3$ for charmonium~\cite{plb612215}.
The asymptotic forms of $\Phi_{asy}(x)$ depend on the corresponding twists for $\chi_{c1}(2P)$.
\end{appendix}


\begin{thebibliography}{99}
\bibitem{pdg2020}
Particle Data Group, Review of particle physics, Prog. Theor. Exp. Phys. \textbf{2020}, 083C01 (2020).
\bibitem{prd74051103}
R.~Kumar \textit{et al.} (Belle Collaboration),
Observation of $B^{\pm}\rightarrow \chi_{c1}\pi^{\pm}$ and search for direct $CP$ violation,
Phys. Rev. D \textbf{74}, 051103 (2006).
\bibitem{prd78091104}
R. Kumar {\it et al.} (Belle Collaboration), Evidence for $B^{0}\rightarrow \chi_{c1}\pi^{0}$ at Belle, Phys. Rev. D \textbf{78}, 091104 (2008).
\bibitem{plb634155}
N. Soni  {\it et al.} (Belle Collaboration), Measurement of branching fractions for $B\rightarrow \chi_{(c1(2))}K(K^{*})$ at BELLE, Phys. Lett. B \textbf{634}, 155 (2006).
\bibitem{prl89011803}
K. Abe {\it et al.} (Belle Collaboration), Observation of $\chi_{(c2)}$ Production in B Meson Decay, Phys. Rev. Lett. \textbf{89}, 011803 (2002).

\bibitem{prl94141801}
B. Aubert {\it et al.} ($BABAR$ Collaboration), Measurement of Branching Fractions and Charge Asymmetries for Exclusive B Decays to Charmonium, Phys. Rev. Lett. \textbf{94}, 141801 (2005).
\bibitem{prd78072004}
R. Mizuk {\it et al.} (Belle Collaboration), Observation of two resonancelike structures in the $\pi^+\chi_{c1}$ mass distribution
in exclusive $\bar{B}^0\rightarrow K^-\pi^+\chi_{c1}$ decays, Phys. Rev. D \textbf{78}, 072004 (2008).
\bibitem{jhep081912018}
R. Aaij {\it et al.} (LHCb Collaboration), Observation of the decay  $\bar{B}_s^0\rightarrow \chi_{c2}K^+K^-$ in the $\phi$  mass region,
J. High Energy Phys. 08 (2018) 191.
\bibitem{prd93052016}
V. Bhardwaj {\it et al.} (Belle Collaboration),
Inclusive and exclusive measurements of B decays to $\chi_{c1}$ and $\chi_{c2}$ at Belle, Phys. Rev. D \textbf{93}, 052016 (2016).


\bibitem{prl91262001}
S. Choi {\it et al.} (Belle Collaboration), Observation of a Narrow Charmoniumlike State in Exclusive $B^{\pm} \to K^{\pm}\pi^{+}\pi^{-}J/\psi$ Decays, Phys. Rev. Lett. \textbf{91}, 262001 (2003).
\bibitem{prl93072001}
D. Acosta {\it et al.} (CDF Collaboration), Observation of the Narrow State $X(3872) \to J/\psi \pi^+ \pi^-$ in $\bar{p}p$ Collisions at $\sqrt{s} = 1.96$ TeV, Phys. Rev. Lett. \textbf{93}, 072001 (2004).
\bibitem{D0:2004zmu}
V. Abazov {\it et al.} (D0 Collaboration), Observation and Properties of the $X(3872)$ Decaying to $J/\psi \pi^+ \pi^-$ in $p\bar{p}$ Collisions at $\sqrt{s} = 1.96$ TeV, Phys. Rev. Lett. \textbf{93}, 162002 (2004).
\bibitem{BaBar:2004oro}
B. Aubert {\it et al.} ($BABAR$ Collaboration), Study of the $B\to J/\psi K^- \pi^+ \pi^-$ decay and measurement of the $B \to X(3872) K^-$ branching fraction, Phys. Rev. D \textbf{71}, 071103 (2005).
\bibitem{LHCb:2011zzp}
R. Aaij {\it et al.} (LHCb Collaboration), Observation of $X(3872) $ production in $pp$ collisions at $\sqrt{s}=7$ TeV, Eur. Phys. J. C \textbf{72}, 1972 (2012).
\bibitem{prd91051101}
A. Bala {\it et al.} (Belle Collaboration),
Observation of $X(3872)$ in $B\rightarrow X(3872)K\pi$ decays, Phys. Rev. D \textbf{91}, 051101 (2015).
\bibitem{prd84052004}
S.-K. Choi {\it et al.} (Belle Collaboration),
Bounds on the width, mass difference and other properties of $X(3872) \to \pi^+ \pi^- J/\psi$ decays, Phys. Rev. D \textbf{84}, 052004 (2011).




\bibitem{prl98132002}
A. Abulencia {\it et al.} (CDF Collaboration), Analysis of the
Quantum Numbers $J^{PC}$ of the $X(3872)$, Phys. Rev. Lett. \textbf{98}, 132002 (2007).

\bibitem{prl110222001}
R. Aaij {\it et al.} (LHCb Collaboration), Determination of the
$X(3872)$ Meson Quantum Numbers, Phys. Rev. Lett. \textbf{110}, 222001 (2013).
\bibitem{prd92011102}
R. Aaij {\it et al.} (LHCb Collaboration), Quantum numbers of the $X(3872)$ state and orbital angular momentum in its
$\rho^0 J/\psi$ decay, Phys. Rev. D \textbf{92}, 011102 (2015).
\bibitem{jhep042013154}
S. Chatrchyan {\it et al.} (CMS Collaboration), Measurement of the $X$(3872) production cross section via decays to $J/\psi \pi^+ \pi^-$ in $pp$ collisions at $\sqrt{s}$ = 7 TeV, J. High Energy Phys. 04 (2013) 154.


\bibitem{prl96102002}
A. Abulencia {\it et al.}  (CDF Collaboration), Measurement of
the Dipion Mass Spectrum in $X(3872)\rightarrow J/\psi \pi^+\pi^-$ Decays, Phys. Rev. Lett. \textbf{96}, 102002 (2006).

\bibitem{jhep012017117}
ATLAS Collaboration, Measurements of $\psi(2S)$ and
$X(3872)\rightarrow J/\psi \pi^+\pi^-$ production in pp collisions at $\sqrt{s}=8$  TeV with the ATLAS detector,
J. High Energy Phys. 01 (2017) 117.
\bibitem{prl126092001}
R. Aaij {\it et al.} (LHCb Collaboration), Observation of Multiplicity Dependent Prompt $\chi_{c1}(3872)$ and $\psi(2S)$ Production in $pp$ Collisions, Phys. Rev. Lett. \textbf{126}, 092001 (2021).
\bibitem{prd102092005}
R. Aaij {\it et al.} (LHCb Collaboration), Study of the line shape of the $\chi_{c1}(3872)$ state, Phys. Rev. D \textbf{102}, 092005 (2020).
\bibitem{jhep082020123}
R. Aaij {\it et al.} (LHCb Collaboration), Study of the $\psi_2(3823)$ and $\chi_{c1}(3872)$ states in $B^+ \rightarrow \left( J\psi\pi^+\pi^-\right)K^+$ decays, J. High Energy Phys. 08 (2020) 123.




\bibitem{explanations}
N.N. Achasov and E.V. Rogozina, $X(3872)$, $I^G(J^{PC}) = 0^+(1^{++})$, as the $\chi_{c1}(2P)$
charmonium, Mod. Phys. Lett. A \textbf{30}, 1550181 (2015). 

\bibitem{Tornqvist:2004qy}
N.A. Tornqvist, Isospin breaking of the narrow charmonium state of Belle at 3872 MeV as a deuson, Phys. Lett. B \textbf{590}, 209 (2004).

\bibitem{Swanson:2003tb}
E.S. Swanson, Short range structure in the $X(3872)$, Phys. Lett. B \textbf{588}, 189 (2004).
\bibitem{Wong:2003xk}
C.-Y. Wong,  Molecular states of heavy quark mesons, Phys. Rev. C \textbf{69}, 055202 (2004). 
\bibitem{Maiani:2004vq}
L.~Maiani, F.~Piccinini, A.~D.~Polosa, and V.~Riquer,
Diquark-antidiquarks with hidden or open charm and the nature of X(3872),
Phys. Rev. D \textbf{71}, 014028 (2005).
\bibitem{Li:2004sta}
B.A. Li, Is $X(3872)$ a possible candidate of hybrid meson?, Phys. Lett. B \textbf{605}, 306 (2005). 
\bibitem{Seth:2004zb}
K.K. Seth, An alternative interpretation of $X(3872)$, Phys. Lett. B \textbf{612},1 (2005). 
\bibitem{Matheus:2009vq}
R.~D.~Matheus, F.~S.~Navarra, M.~Nielsen, and C.~M.~Zanetti,
QCD sum rules for the X(3872) as a mixed molecule-charmoniun state,
Phys. Rev. D \textbf{80}, 056002 (2009).



\bibitem{Suzuki:2005ha}
M.~Suzuki,
The X(3872) boson:  Molecule or charmonium,
Phys. Rev. D \textbf{72}, 114013 (2005).


\bibitem{Kalashnikova:2005ui}
Y.~S.~Kalashnikova, Coupled-channel model for charmonium levels and an option for $X(3872)$, Phys. Rev. D \textbf{72}, 034010 (2005).


\bibitem{Takizawa:2012hy}
M.~Takizawa and S.~Takeuchi, $X(3872)$ as a hybrid state of charmonium and the hadronic molecule, Prog. Theor. Exp. Phys. \textbf{2013}, 093D01 (2013).


\bibitem{Chen:2013pya}
W.~Chen, H.~y.~Jin, R.~T.~Kleiv, T.~G.~Steele, M.~Wang, and Q.~Xu,
QCD sum-rule interpretation of X(3872) with $J^{PC}=1^{++}$ mixtures of hybrid charmonium and $\overline{D}D^*$ molecular currents,
Phys. Rev. D \textbf{88}, 045027 (2013).



\bibitem{Wallbott:2019dng}
P.~C.~Wallbott, G.~Eichmann, and C.~S.~Fischer,
$X(3872)$ as a four-quark state in a Dyson-Schwinger/Bethe-Salpeter approach,
Phys. Rev. D \textbf{100}, 014033 (2019).

\bibitem{Matheus:2006xi}
R.~D.~Matheus, S.~Narison, M.~Nielsen, and J.~M.~Richard,
Can the X(3872) be a $1^{++}$ four-quark state?,
Phys. Rev. D \textbf{75}, 014005 (2007).

\bibitem{Dubnicka:2010kz}
S.~Dubnicka, A.~Z.~Dubnickova, M.~A.~Ivanov, and J.~G.~Korner,
Quark model description of the tetraquark state X(3872) in a relativistic constituent quark model with infrared confinement,
Phys. Rev. D \textbf{81}, 114007 (2010).

\bibitem{Colangelo:2007ph}
P. Colangelo, F. De Fazio, and S. Nicotri, $X(3872) \to D \bar{D} \gamma$ decays and the structure of $X(3872)$, Phys. Lett. B \textbf{650}, 166 (2007).



\bibitem{Butenschoen:2019npa}
M. Butenschoen, Z.-G. He, and B. A. Kniehl, Deciphering the $X(3872)$ Via Its Polarization in Prompt Production at the CERN LHC, Phys. Rev. Lett. \textbf{123}, 032001 (2019).
\bibitem{Coito:2012vf}
S. Coito, G. Rupp, and E. van Beveren, $X(3872)$ is not a true molecule, Eur. Phys. J. C \textbf{73}, 2351 (2013).







\bibitem{rmp90015003}
S. L. Olsen, T. Skwarnicki, and D. Zieminska, Nonstandard heavy mesons and baryons: Experimental evidence,
Rev. Mod. Phys. \textbf{90}, 015003 (2018).
\bibitem{plb59191}
Bla$\check{z}$enka  Meli$\acute{c}$, LCSR analysis of exclusive two body $B$ decay into charmonium, Phys. Lett. B \textbf{591}, 91 (2004).
\bibitem{plb568127}
Z.-Z. Song and K.-T. Chao, Problems of QCD factorization in exclusive decays of $B$ meson to charmonium, Phys. Lett. B \textbf{568}, 127 (2003).
\bibitem{prd69054009}
Z.-Z. Song, C. Meng, Y.-J. Gao, and K.-T. Chao, Infrared divergences of $B$ meson exclusive decays to $P$ wave charmonia in QCD factorization and nonrelativistic QCD, Phys. Rev. D \textbf{69}, 054009 (2004).
\bibitem{prd59054003}
M. Beneke, F. Maltoni, and I. Z. Rothstein, QCD analysis of inclusive $B$ decay into charmonium, Phys. Rev. D \textbf{59}, 054003 (1999).
\bibitem{npb811155}
M. Beneke and L. Vernazza, $B \to \chi_{cJ} K$ decays revisited, Nucl. Phys. \textbf{B811}, 155 (2009).
 \bibitem{prd71114008}
C.-H. Chen and H.-N. Li, Nonfactorizable contributions to B meson decays into charmonia, Phys. Rev. D \textbf{71}, 114008 (2005).
\bibitem{epjc78463}
Z. Rui, Q. Zhao, and L. Zhang, Branching ratios and $CP$ asymmetries of $B\rightarrow \chi_{c1}K(\pi)$ decays, Eur. Phys. J. C \textbf{78}, 463 (2018).
\bibitem{prd87074035}
C. Meng, Y.J. Gao, K.T. Chao, $B \rightarrow \chi_{c1}(1P,2P)K$ decays in QCD factorization and $X(3872)$, Phys. Rev. D \textbf{87}, 074035 (2013).
\bibitem{epjc49643}
X. Liu and Y. M. Wang, Revisiting $B^{+}\rightarrow X(3872)+ K^{+}$ in pQCD assigning to $X(3872) 2^{3}P_{1}$ charmonium, Eur. Phys. J. C \textbf{49}, 643 (2007).
\bibitem{Wang:2007fs}
Y.~M.~Wang and C.~D.~Lu,
Weak productions of new charmonium in semileptonic decays of $B_c$,
Phys. Rev. D \textbf{77}, 054003 (2008).


\bibitem{prd97012005}
Y. Kato {\it et al.} (Belle Collaboration),  Measurements of the absolute branching fractions
of $B^+\rightarrow X_{c\bar c} K^+$  and $B^+\rightarrow \bar {D}^{(*)0\pi^{+}}$  at Belle, Phys. Rev. D \textbf{97}, 012005 (2018).
\bibitem{Nakamura:2019emd}
S.~X.~Nakamura,
Triangle singularities in $\bar{B}^0\to \chi_{c1}K^-\pi^+$ relevant to $Z_1(4050)$ and $Z_2(4250)$,
Phys. Rev. D \textbf{100}, 011504(R)(2019).
\bibitem{Nakamura:2019btl}
S.~X.~Nakamura and K.~Tsushima,
$Z_c(4430)$ and $Z_c(4200)$ as triangle singularities,
Phys. Rev. D \textbf{100}, 051502(R) (2019).
\bibitem{Nakamura:2019nch}
S.~X.~Nakamura,
$Z_c(4430)$, $Z_c(4200)$, $Z_1(4050)$, and $Z_2(4250)$ as triangle singularities,
AIP Conf. Proc. \textbf{2249}, 030006 (2020).



\bibitem{prd91094024}
W.~F.~Wang, H.~n.~Li, W.~Wang, and C.~D.~L\"u,
$S$-wave resonance contributions to the $B^0_{(s)}\to J/\psi\pi^+\pi^-$ and $B_s\to\pi^+\pi^-\mu^+\mu^-$ decays,
Phys. Rev. D \textbf{91}, 094024 (2015).
\bibitem{prd97033006}
Z. Rui, and W.~F.~Wang, $S$-wave $K\pi$ contributions to the hadronic charmonium $B$ decays in the perturbative QCD approach, Phys. Rev. D \textbf{97}, 033006 (2018).
\bibitem{prd98113003}
Z. Rui, Y. Li, and H. N. Li, $P$-wave contributions to $B\to\psi\pi\pi$ decays in perturbative QCD approach, Phys. Rev. D \textbf{98}, 113003 (2018).
\bibitem{prd99093007}
Z. Rui, Y. Q. Li,  and J. Zhang, Isovector scalar $a_0(980)$ and $a_0(1450)$ resonances in the $B\rightarrow \psi (K\bar{K},\pi\eta)$ decays, Phys. Rev. D \textbf{99}, 093007 (2019).
\bibitem{epjc79792}
Z. Rui, Y. Li, and Hong Li, Studies of the resonance components in the $B_s$ decays into charmonia plus kaon pair, Eur. Phys. J. C \textbf{79}, 792 (2019).
\bibitem{prd101016015}
Y. Li , D.-C. Yan,  Z. Rui, and Z.-J. Xiao, $S$, $P$ and $D$-wave resonance contributions to $B_{(s)} \to \eta_c(1S,2S) K\pi$ decays in the perturbative QCD approach, Phys. Rev. D \textbf{101}, 016015 (2020).
\bibitem{cpc44073102}
Y. Li ,   Z. Rui, and  Z.-J. Xiao, $P$-wave contributions to $B_{(s)}\to\psi K\pi$ decays in perturbative QCD approach, Chin. Phys. C \textbf{44}, 073102 (2020).

\bibitem{npb924745}
Y. Li , A.-J. Ma,  Z. Rui, and  Z.-J. Xiao, Quasi-two-body decays $B \to \eta_c {(1S ,2S)}\;[\rho(770),\rho(1450),\rho(1700) \to ]\; \pi\pi$ in the perturbative QCD approach, Nuc. Phys. \textbf{B924}, 745 (2017).

\bibitem{zjhep}
Z. Rui, Y. Li, and H.~n.~Li, Four-body decays $B_{(s)} \rightarrow (K\pi)_{S/P} (K\pi)_{S/P}$ in the perturbative QCD approach,
 J. High Energy Phys. 05 (2021) 082.

\bibitem{Li:2021qiw}
Y.~Li, D.~C.~Yan, Z.~Rui, and Z.~J.~Xiao,
Study of $B_{(s)} \to (\pi\pi)(K\pi)$ decays in the perturbative QCD approach,
arXiv:2107.10684.



\bibitem{G} A. G. Grozin,
On wave functions of meson pairs and meson resonances, Sov. J. Nucl. Phys. {\bf 38}, 289 (1983).

\bibitem{G1}
A. G. Grozin, One- and two-particle wave functions of multihadron systems, Theor. Math. Phys. {\bf 69}, 1109 (1986).


\bibitem{DM}
D. M\"uller,  D. Robaschik, B. Geyer,  F.-M. Dittes, and  J.~Ho\v rej\v si,
 Wave functions, evolution equations and evolution kernels from light ray operators of QCD,
 Fortschr. Phys. {\bf 42}, 101 (1994).

\bibitem{Diehl:1998dk}
  M.~Diehl, T.~Gousset, B.~Pire, and O.~Teryaev,
 Probing Partonic Structure in $\gamma^{*}\gamma \rightarrow \pi\pi$ Near Threshold,
  Phys.\ Rev.\ Lett.\ {\bf 81}, 1782 (1998).
\bibitem{Diehl:1998dk1}
  M.~Diehl, T.~Gousset, and B.~Pire,
Exclusive production of pion pairs in $\gamma^{*}\gamma$ collisions at large $Q^{2}$, Phys. Rev. D {\bf 62}, 073014 (2000) .
  \bibitem{Diehl:1998dk2}
B.~Pire and L.~Szymanowski, Impact representation of generalized distribution amplitudes, Phys. Lett. B {\bf556}, 129 (2003).
\bibitem{MP}
M.V. Polyakov, Hard exclusive electroproduction of two pions and their resonances,
Nucl. Phys. {\bf B555}, 231 (1999).
\bibitem{210503899}
Y. Li, D. C. Yan, J. Hua, Z. Rui, and H. n. Li,
Global determination of two-meson distribution amplitudes from three-body $B$ decays in the perturbative QCD approach, arXiv: 2105.03899.


\bibitem{ppnp5185}
H.~n.~Li, QCD aspects of exclusive $B$ meson decays, Prog. Part. Nucl. Phys. \textbf{51}, 85 (2003), and references therein.


\bibitem{Li:2014xda}
H.~n.~Li and Y.~M.~Wang,
Non-dipolar Wilson links for transverse-momentum-dependent wave functions, J. High Energy Phys. 06 (2015) 013.
\bibitem{Li:2012nk}
H.~n.~Li, Y.~L.~Shen, and Y.~M.~Wang,
Next-to-leading-order corrections to $B \to \pi$ form factors in $k_T$ factorization, Phys. Rev. D \textbf{85}, 074004 (2012).
 \bibitem{prd65014007}
T. Kurimoto, H.~n.~Li, and A. I. Sanda, Leading-power contributions to $B\rightarrow\pi, \rho$ transition form factors, Phys. Rev. D \textbf{65}, 014007 (2001).
\bibitem{201215074}
J. Hua, H.~n.~Li, C. D. L\"{u}, W. Wang, and Zhi-Peng Xing,
Global analysis of hadronic two-body $B$ decays in the perturbative QCD approach, Phys. Rev. D \textbf{104}, 016025 (2021).
\bibitem{prd102011502}
W. Wang, Y. M. Wang, J. Xu, and S. Zhao, $B$-meson light-cone distribution amplitude from the Euclidean quantity,
Phys. Rev. D \textbf{102}, 011502(R) (2020).
\bibitem{prd70074030}
H.~n.~Li  and H. S. Liao, $B$ meson wave function in $k_T$ factorization, Phys. Rev. D \textbf{70}, 074030 (2004).
\bibitem{Li:2012md}
H.~n.~Li, Y.~L.~Shen, and Y.~M.~Wang,
Resummation of rapidity logarithms in $B$ meson wave functions,
J. High Energy Phys. 02 (2013) 008.
\bibitem{prd97033001}
Z. Rui, Probing the $P$-wave charmonium decays of $B_c$ meson, Phys. Rev. D \textbf{97}, 033001 (2018).

\bibitem{Wang:2013ywc}
X.~P.~Wang and D.~Yang,
The leading twist light-cone distribution amplitudes for the S-wave and P-wave quarkonia and their applications in single quarkonium exclusive productions, J. High Energy Phys. 06 (2014) 121.

\bibitem{Li:2020app}
Y.~Q.~Li, M.~K.~Jia, and Z.~Rui,
Revisiting nonfactorizable contributions to factorization-forbidden decays of $B$ mesons to charmonium,
Chin. Phys. C \textbf{44}, 113104 (2020).



\bibitem{plb561258}
 C. H. Chen and H. N. Li, Three body nonleptonic $B$ decays in perturbative QCD, Phys. Lett. B \textbf{561}, 258 (2003).

\bibitem{prd73014017}
H.-Y. Cheng, C.-K. Chua, and K.-C. Yang, Charmless hadronic $B$ decays involving scalar mesons: Implications on the nature
of light scalar mesons, Phys. Rev. D \textbf{73}, 014017 (2006).
\bibitem{prd77014034}
H.-Y. Cheng, C.-K. Chua, and K.-C. Yang, Charmless $B$ decays to a scalar meson and a vector meson, Phys. Rev. D \textbf{77}, 014034 (2008).
\bibitem{Stone:2013eaa}
S.~Stone and L.~Zhang,
Use of $B\to J/\psi f_0$ Decays to Discern the $q \bar{q}$ or Tetraquark Nature of Scalar Mesons, Phys. Rev. Lett. \textbf{111}, 062001 (2013).


\bibitem{Fleischer:2011au}
R.~Fleischer, R.~Knegjens, and G.~Ricciardi, Anatomy of $B^0_{s,d} \to J/\psi f_0(980)$, Eur. Phys. J. C \textbf{71}, 1832 (2011).
\bibitem{Jaffe:2004ph}
R.~L.~Jaffe, Exotica, Phys. Rep. \textbf{409}, 1 (2005).


\bibitem{Klempt:2007cp}
E.~Klempt and A.~Zaitsev,
Glueballs, hybrids, multiquarks. Experimental facts versus QCD inspired concepts,
Phys. Rept. \textbf{454}, 1 (2007).
\bibitem{Aaij:2014emv}
R.~Aaij \textit{et al.} (LHCb Collaboration),
Measurement of resonant and $CP$ components in $\bar{B}_s^0\to J/\psi\pi^+\pi^-$ decays,
Phys. Rev. D \textbf{89}, 092006 (2014).
\bibitem{cpc41083105}
A.-J. Ma, Y. Li, W. F. Wang, and Z.-J. Xiao, $S$-wave resonance contributions to the $B^0_{(s)}\to \eta_c{(2S)}\pi^+\pi^-$ in the perturbative QCD factorization approach,  Chin. Phys. C \textbf{41}, 083105 (2017).
\bibitem{epjc76675}
Y. Li, A.-J. Ma, W. F. Wang, and Z.-J. Xiao,  The S-wave resonance contributions to the three-body decays $B^0_{(s)}\rightarrow \eta _c f_0(X)\rightarrow \eta _c\pi ^+\pi ^-$ in perturbative QCD approach, Eur. Phys. J. C \textbf{76}, 675 (2016).
\bibitem{epjc77199}
Z. Rui, Y. Li, and W. F. Wang, The S-wave resonance contributions in the $B^0_s$ decays into $ \psi(2S,3S)$ plus pion pair, Eur. Phys. J. C \textbf{77}, 199 (2017).
\bibitem{plb63228}
S. M. Flatt\'{e}, On the nature of $0^{+}$ mesons, Phys. Lett. \textbf{63B}, 228 (1976).

\bibitem{jpg34151}
D.~V.~Bugg, The mass of the sigma pole, J. Phys. G \textbf{34}, 151 (2007).
\bibitem{Aaij:2015sqa}
R.~Aaij \textit{et al.} (LHCb Collaboration),
Dalitz plot analysis of $B^0 \to \overline{D}^0 \pi^+\pi^-$ decays,
Phys. Rev. D \textbf{92}, 032002 (2015).


\bibitem{Aaij:2014vda}
R.~Aaij \textit{et al.} (LHCb Collaboration),
Measurement of the $CP$-violating phase $\beta$ in $B^0\rightarrow J/\psi \pi^+\pi^-$ decays and limits on penguin effects,
Phys. Lett. B \textbf{742}, 38 (2015).
\bibitem{prd90012003}
R.~Aaij \textit{et al.} (LHCb Collaboration), Measurement of the resonant and CP components in $\overline{B}^0\to J/\psi \pi^+\pi^-$ decays, Phys. Rev. D \textbf{90}, 012003 (2014).
\bibitem{Pelaez:2015qba}
J.~R.~Pelaez,
From controversy to precision on the sigma meson: A review on the status of the non-ordinary $f_0(500)$ resonance,
Phys. Rep. \textbf{658}, 1 (2016).

\bibitem{Oller:2004xm}
J.~A.~Oller,
Final state interactions in hadronic D decays,
Phys. Rev. D \textbf{71}, 054030 (2005).


\bibitem{LHCb:2019sus}
R.~Aaij \textit{et al.} (LHCb Collaboration),
Amplitude analysis of the $B^+ \rightarrow \pi^+\pi^+\pi^-$ decay,
Phys. Rev. D \textbf{101}, 012006 (2020).


\bibitem{Cheng:2020iwk}
H.~Y.~Cheng, C.~W.~Chiang, and C.~K.~Chua,
Finite-width effects in three-Body B decays,
Phys. Rev. D \textbf{103}, 036017 (2021).


\bibitem{npb296493}
D. Aston {\it et al.} (LASS Collaboration), A study of $K^-\pi^+$ scattering in the reaction $K^-p\rightarrow K^-\pi^+$ at 11GeV/$c$,
 Nucl. Phys. \textbf{B296}, 493 (1988).
 \bibitem{Aubert:2005ce}
B.~Aubert \textit{et al.} ($BABAR$ Collaboration),
Dalitz-plot analysis of the decays $B^\pm \to K^\pm \pi^\mp \pi^\pm$,
Phys. Rev. D \textbf{72}, 072003 (2005).
\bibitem{BaBar:2005qms}
B.~Aubert \textit{et al.} ($BABAR$ Collaboration),
Dalitz-plot analysis of the decays $B^\pm \to K^\pm \pi^\mp \pi^\pm$,
 Phys. Rev. D \textbf{74}, 099903 (2006).

\bibitem{Bugg:2008ig}
D.~V.~Bugg,
Reanalysis of data on $a_0(1450)$  and $a_0(980)$,
Phys. Rev. D \textbf{78}, 074023 (2008).

\bibitem{epjc77610}
Z. Rui, Y. Li, and Zhen-Jun Xiao, Branching ratios, $CP$ asymmetries and polarizations of $B\rightarrow \psi(2S) V$ decays, Eur. Phys. J. C \textbf{77}, 610 (2017).
\bibitem{epjc11695}
H.-N. Li and B. Melic, Determination of heavy meson wave functions from B decays, Eur. Phys. J. C \textbf{ 11 }, 695 (1999).

\bibitem{Liu:2020upy}
  X.~Liu, H.~n.~Li and Z.~J.~Xiao,
 Next-to-leading-logarithm $k_T$ resummation for $B_c\to J/\psi$ decays,
  Phys. Lett. B \textbf{811}, 135892 (2020).


\bibitem{qcdf}
M. Beneke, G. Buchalla, M. Neubert, and C.T. Sachrajda, QCD Factorization for $B \rightarrow \pi\pi$ Decays: Strong Phases and $CP$ Violation in the Heavy Quark Limit, Phys. Rev. Lett. \textbf{83}, 1914 (1999).
 \bibitem{Beneke:2000ry}
M.~Beneke, G.~Buchalla, M.~Neubert, and C.~T.~Sachrajda,
QCD factorization for exclusive, nonleptonic B meson decays: General arguments and the case of heavy light final states, Nucl. Phys. \textbf{B591}, 313 (2000).
 \bibitem{Beneke:2001ev}
M.~Beneke, G.~Buchalla, M.~Neubert, and C.~T.~Sachrajda,
QCD factorization in $B\rightarrow \pi K, \pi\pi$ decays and extraction of Wolfenstein parameters, Nucl. Phys. \textbf{B606}, 245 (2001).

\bibitem{BaBar:2008bxw}
B.~Aubert \textit{et al.} ($BABAR$ Collaboration),
Search for the $Z(4430)^{-}$ at $BABAR$, Phys. Rev. D \textbf{79}, 112001 (2009).
\bibitem{Lees:2011ik}
J.~P.~Lees \textit{et al.} ($BABAR$ Collaboration),
Search for the $Z_1(4050)^+$ and $Z_2(4250)^+$ states in $\bar B^0 \to \chi_{c1} K^- \pi^+$ and $B^+ \to \chi_{c1} K^0_S \pi^+$,
Phys. Rev. D \textbf{85}, 052003  (2012).
\bibitem{Aubert:2006nu}
B.~Aubert \textit{et al.} ($BABAR$ Collaboration),
Dalitz plot analysis of the decay $B^\pm \to K^\pm K^\pm K^\mp$, Phys. Rev. D \textbf{74}, 032003 (2006).

\bibitem{Ablikim:2005kp}
M.~Ablikim \textit{et al.} (BES Collaboration),
Partial wave analysis of $\chi_{c0} \rightarrow \pi^{+}\pi^{-} K^{+} K^{-}$,
Phys. Rev. D \textbf{72}, 092002 (2005).
\bibitem{plb612215}
A. E. Bondar and V. L. Chernyak, Is the BELLE result for the cross section $\sigma(e^{+} e^{-}\rightarrow J /\psi+ \eta_{c})$ a real difficulty for QCD? , Phys. Lett. B \textbf{612}, 215 (2005).
\end{thebibliography}
\end{document}